\documentclass[]{pasj01}
\Received{}%{yyyy/mm/dd}
\Accepted{}%{yyyy/mm/dd}
\let\uproot\relax
\let\leftroot\relax

\usepackage{amsmath}
\usepackage{hyphenat} % allow break
\usepackage{url}
\usepackage{bm}
\usepackage{multirow}
\usepackage[switch,mathlines]{lineno}
\usepackage{comment}
\usepackage{lscape}

\begin{document}

\title{ 
%\LETTERLABEL %%% <-- uncomment for LETTER article  
%\REVIEWLABEL %%% <-- uncomment for REVIEW article  
HINOTORI I: The Nature of Rejuvenation Galaxies}

%%% begin:list of authors
% Do NOT capitalize all letters in "textsc".
\author{Takumi S. \textsc{Tanaka}
\altaffilmark{1,2,3}$^\ast$}
\author{Kazuhiro \textsc{Shimasaku}
\altaffilmark{1,4}}
\author{Sandro \textsc{Tacchella}
\altaffilmark{5,6}}
\author{Makoto \textsc{Ando}
\altaffilmark{1}}
\author{Kei \textsc{Ito}
\altaffilmark{1}}
\author{Hassen M. \textsc{Yesuf}
\altaffilmark{2,7}}
\author{Suin \textsc{Matsui}
\altaffilmark{1}}

\altaffiltext{1}{Department of Astronomy, Graduate School of Science, The University of Tokyo, 7-3-1 Hongo, Bunkyo-ku, Tokyo 113-0033, Japan}

\altaffiltext{2}{Kavli Institute for the Physics and Mathematics of the Universe, The University of Tokyo, Kashiwa, Chiba 277-8583, Japan}

\altaffiltext{3}{Center for Data-Driven Discovery, Kavli IPMU (WPI), UTIAS, The University of Tokyo, Kashiwa, Chiba 277-8583, Japan}

\altaffiltext{4}{Research Center for the Early Universe, Graduate School of Science, The University of Tokyo, 7-3-1 Hongo, Bunkyo-ku, Tokyo 113-0033, Japan}

\altaffiltext{5}{Kavli Institute for Cosmology, University of Cambridge, Madingley Road, Cambridge CB3 0HA, UK}

\altaffiltext{6}{Cavendish Laboratory, University of Cambridge, 19 JJ Thomson Avenue, Cambridge CB3 0HE, UK}

\altaffiltext{7}{Kavli Institute for Astronomy and Astrophysics, Peking University, Beijing 100871, People's Republic of China}

\email{takumi.tanaka@ipmu.jp}

%%% end:list of authors

%% `\KeyWords{}' always has to be placed before ``\maketitle'' 
%%  List of Key Words:  https://academic.oup.com/pasj/pages/Pasj_Keywords 
\KeyWords{galaxies: evolution --- galaxies: star formation --- galaxies: stellar content}

\maketitle

\begin{abstract}
We present the {\sc Hinotori} (star formation History INvestigatiOn TO find RejuvenatIon) project to reveal the nature of rejuvenation galaxies (RGs), which are galaxies that restarted their star formation after being quiescent.
As the first step of {\sc Hinotori}, we construct the largest RG sample with 1071 sources.
We select these RGs from 8857 MaNGA (Mapping Nearby Galaxies at APO) survey galaxies by reconstructing their star formation histories with {\sc Prospector} spectral energy distribution fitting code.
Both optical spectral data and UV to IR photometric data are used for the fitting.
Using mock data, we confirm that our method can detect weak rejuvenation events that form only about 0.1\% of the total stellar mass with high completeness.
The RGs account for $\sim10\%$ of the whole sample, and rejuvenation events contribute on average only about 0.1\% of the total stellar mass in those galaxies but 17\% of the cosmic-star formation rate density today.
Our RGs have a similar mass distribution to quiescent galaxies (QGs).
However, the morphology of the RGs is more disk-like than QGs, suggesting that rejuvenation may occur selectively in disk-like QGs.
Our results also suggest the possibility of multiple-time rejuvenation events in a single galaxy.
Further spatially resolved analyses of integral field unit data and radio observations and comparisons to simulations are needed to identify the mechanism and the role of rejuvenation in galaxy evolution.
\end{abstract}
%\linenumbers
%\pagewiselinenumbers

%%%%%%%%%%%%%%%%% Introduction %%%%%%%%%%%%%%%%%%
\section{Introduction}\label{s1}
Galaxies segregate into two distinct groups in the star-formation rate (SFR) - stellar mass ($M^*$) diagram: ``star-forming galaxies'' (SFGs) with active star formation and ``quiescent galaxies'' (QGs) with low-to-no star formation (e.g., \cite{Renzini2015, Feldmann2017}).
Because the fraction of QGs increases with cosmic time (e.g., \cite{Ilbert2013, Muzzin2013, Davidzon2017, Weaver2022}), a widely-accepted simple galaxy evolution scenario is that SFGs stop star formation (quenching) and become QGs (e.g., \cite{Faber2007, Peng2010}).
Galaxies in between SFGs and QGs are called ``green valley'' galaxies (GVs).
GVs are widely thought to be galaxies in transition from SFGs to QGs (e.g., \cite{Salim2014,Angthopo2020}).

Recent progress in the spectral energy distribution (SED) fitting technique (\cite{Walcher2011} for review) has enabled us to search for and analyze galaxies with unique star-formation histories (SFHs), such as ``rejuvenating galaxies", which have been overlooked in previous studies. Rejuvenating galaxies have resumed their star formation after being quiescent (e.g., \cite{Trayford2016,Cleland2021,Zhang2022}).
We may need to update the simple evolution scenario from SFGs to QGs to explain rejuvenating galaxies.

Some studies (e.g., \cite{Chauke2019,Mancini2019,Tacchella2022}) have also focused on ``rejuvenated galaxies''.
The difference between ``rejuvenating galaxies'' and ``rejuvenated galaxies'' is that the latter have terminated their resumed star-formation and have been quiescent again.
Since both populations have experienced rejuvenation events, this paper deals with both populations and collectively refers to them as ``rejuvenation galaxies'' (RGs).
We do not use the abbreviation ``RG'' when distinguishing between ``rejuvenated galaxies'' and ``rejuvenating galaxies''.

RGs have been studied since the 2000s. 
Early studies, which are limited to the low-$z$ universe, have selected early-type galaxies (ETGs) with recent star formation as RGs by using UV detection (\cite{Kaviraj2007,Donas2007,Schawinski2007}), spectral and photometric features (\cite{Treu2005,Cleland2021}), or a stellar population synthesis (\cite{Thomas2010}).
The fraction of RGs (or ETGs with recent star formation) in previous samples ranges from $10\%$ to $30\%$ depending on the selection method and the properties of the parent samples.
Recent developments in SED fitting methods and spectroscopic surveys in the intermediate-$z$ universe enabled selecting $0.6\lesssim z\lesssim 2$ RGs based on their SFHs or stellar ages reconstructed by SED fitting (\cite{Belli2017,Gobat2017,Carnall2019,Chauke2019,Mancini2019,Akhshik2021,Tacchella2022,Paspaliaris2023}).

{\bf The mechanism and cause of rejuvenation} and {\bf the role of rejuvenation in galaxy evolution} are major open questions. On the cause of rejuvenation, some studies have suggested that rejuvenation is related to mergers. For example, 
\citet{Kaviraj2009} have used a simulation to show that a minor merger can explain the recent star formation of ETGs. Other studies have also suggested that gas accretion onto elliptical galaxies after merging may cause rejuvenation and enables rejuvenated ellipticals to acquire a disk and turn into red spirals or S0 galaxies (e.g., \cite{Mapelli2015,Diaz2018,Hao2019,Himansh2022}). 
\citet{Paspaliaris2023} have argued that HI gas, as found in some ETGs by \citet{Thom2012}, could cause rejuvenation.
Note, however, that \citet{Park2022} have suggested that red disk-like QGs cannot be described with the above merger-rejuvenation scenario.
Besides, \citet{Martin-Navarro2022} have reported that many type-I active galactic nuclei (AGN) had experienced rejuvenation, suggesting the relationship between rejuvenation events and AGN activities.

Understanding the rejuvenation mechanism will deepen our knowledge of the star formation activity in a galaxy.
In particular, we can consider RGs as former QGs that could not maintain a quiescent state.
Comparing RGs with QGs enables us to discuss the quenching process, especially the conditions needed to maintain a quiescent state.

{\bf The role of rejuvenation in galaxy evolution} also remains to be evaluated.
\citet{Chauke2019} and \citet{Tacchella2022} have reported that, at most, only about 10\% of the total mass is formed in a rejuvenation event.
However, we can detect only recent rejuvenation events because the time resolution of the reconstructed star formation history of a galaxy decreases with look-back time.
Thus, we may underestimate the contributions of rejuvenation events to the total star-formation activities.
The relationship between RGs and GVs is also relevant to the role of rejuvenation.
GVs are widely thought to be in the quenching phase.
However, this simple view needs to be modified if a significant fraction of GVs are RGs.
Furthermore, \citet{Mancini2019} reported that the bending of the star-forming main sequence (SFMS) in the high-mass end is caused by RGs.

Despite the importance of RGs, previous studies have failed to discuss the nature of RGs statistically.
This is mainly because the previous studies are based on either small samples of RGs or biased parent samples (e.g., RGs are selected from only QGs).
For example, \citet{Tacchella2022} have found that RGs reside in massive dark halos based on nine RGs selected from 161 QGs. 
\citet{Cleland2021} have detected no significant environmental dependence of rejuvenation fraction in a sample of $\simeq 350$ RGs. However, these results need to be confirmed with a larger RG sample and with a more inclusive parent sample not limited to QGs. In addition, different definitions of RGs among the previous studies have also made it difficult to compare their results with each other and to compare them with simulation results (e.g., \cite{Kaviraj2009,Trayford2016,Pandya2017,Nelson2018,Behroozi2019,Alarcon2022}).

We launch the {\sc Hinotori}\footnote{star-formation History INvestigatiOn TO find RejuvenatIon. Hinotori means phoenix in Japanese.} project to solve those problems.
{\sc Hinotori} aims to understand the nature of RGs using well-constrained SFHs.
Selecting RGs with reliable SFHs is the most direct selection method.
As the first step of {\sc Hinotori}, we select RGs from a large ($N\sim10^4$) parent sample of $z\simeq 0$ galaxies covering all morphological types.
To reconstruct the SFH for individual galaxies, we run the Bayesian SED fitting library {\sc Prospector} \citep{Leja2017, Johnson2021} on their spectroscopic and UV to MIR photometric data.
In this paper, we present the most extensive catalog to date of SFH-selected RGs and make the first secure statistical analysis of RGs.

Section~\ref{s2} describes the data and how to make the input data for {\sc Prospector}.
Section~\ref{s3} describes the SED fitting method and the selection method of RGs.
In Sections~\ref{s4} and \ref{s5}, we show the results from a mock sample and real data, respectively.
We discuss the nature of RGs in Section~\ref{s6} and present a conclusion and prospects for future RG studies in Section~\ref{s7}.
Throughout this work, we assume the cosmological parameters from the WMAP-9 \citep{Hinshaw2013}.

%%%%%%%%%%%%%%%%% DATA %%%%%%%%%%%%%%%%%%
\section{Data}\label{s2}
\subsection{MaNGA survey}\label{s21}

We construct a parent sample from the sample of the MaNGA survey \citep{Bundy2015}, an integral field unit survey in the SDSS-IV \citep{Blanton2017}.
This survey was carried out with the SDSS 2.5-meter telescope \citep{Gunn2006} equipped with a fiber feed system \citep{Drory2015} and the Baryon Oscillation Spectroscopic Survey (BOSS) spectrograph \citep{Smee2013} covering 3600 – 10300 \AA\ at $R\sim2200$.
The target galaxies of the MaNGA survey were selected from the NASA-Sloan atlas (NSA {\tt v1\_0\_1}).
See \citet{Yan2016} for the survey design.
As detailed in \citet{Wake2017}, the MaNGA sample consists of the following three samples selected with redshifts, magnitudes, and colors to achieve a uniform number density distribution of $M_i$ and additional targets selected for specific sciences.
\begin{itemize}
    \item {\bf Primary Sample}: Galaxies whose $1.5R_e$ area is covered by the IFU.
    \item {\bf Secondary Sample}: Galaxies whose $2.5R_e$ area is covered by the IFU.
    \item {\bf Color-Enhanced Sample}: Selected to fill in the region on the ${\rm NUV}-i$ versus $M_i$ color-magnitude diagram poorly sampled in the Primary and Secondary Samples. Galaxies in this region include low-luminosity red galaxies, high-luminosity blue galaxies, and GVs. The IFU covers their $1.5R_e$, the same as the Primary Sample. The Primary Sample and the Color-Enhanced Sample are combined as the Primary+ Sample.
    \item {\bf Ancillary Targets}\footnote{\url{https://www.sdss4.org/dr17/manga/manga-target-selection/ancillary-targets/}}: Additionally selected 977 targets, such as luminous AGN, galaxies in voids, pairs and mergers, dwarf galaxies, blight cluster galaxies, and so on.
\end{itemize}
These galaxies are distributed over $5\times10^8\leq M^*/M_\odot \leq 3\times10^{11}$ at $0.01\leq z\leq0.15$.
In this study, we use the IFU data of 11,273 datacubes in the SDSS DR17 \citep{Abdurro'uf2022} reduced with the version 3.1.1 MaNGA Data Reduction Pipeline (DRP: \cite{Law2016,Law2021}), which performs flux calibration, sky subtraction, correction of the Galactic dust attenuation, and coadding individual exposures to make the final 3D datacubes of each galaxy.

Assuming that the RG fraction, the number ratio of RGs to total galaxies, is $f_{\rm RG}\sim 10\%$ as suggested by \citet{Nelson2018, Behroozi2019, Chauke2019,Tacchella2022}, we expect that the entire MaNGA sample includes $N\sim10^3$ RGs, enabling a reliable statistical analysis.
Besides, because all selected RGs have IFU spectroscopic data, we can also analyze the rejuvenation mechanism with spatially resolved characteristics.

\subsection{Removing AGN, mergers, and pair galaxies}\label{s22}
In this study, we do not use galaxies hosting active galactic nuclei (AGN) because {\sc Prospector} has no AGN template other than dust torus emission in IR wavelengths \citep{Leja2018}.
\citet{Comerford2020} published an AGN catalog selected from the 8th MaNGA Product Launch (MPL-8) sample.
However, since this sample ($N=6261$ galaxies) is smaller than the DR17 sample we use, we extend the selection method used by \citet{Comerford2020} to the DR17 MaNGA sample to exclude AGN.
In total, we remove 406 AGN from the whole sample.

Merger and pair galaxies are also excluded using the {\tt MANGA TARGET3} information, a flag for ancillary targets, because it is difficult to create a 1D spectrum of a target galaxy by subtracting the contribution from its companion(s) in the 1D-flattening process described in section \ref{s24}.
We remove 119 galaxies that have either {\tt TARGET3\_PAIR\_SIM}, {\tt TARGET3\_PAIR\_RECENTER}, {\tt TARGET3\_PAIR\_2IFU}, or {\tt TARGET3\_PAIR\_ENLARGE} flag.

\subsection{Crossmatching with photometric catalogs}\label{s23}
\begin{figure*}
 \begin{center}
  \includegraphics[width=16cm]{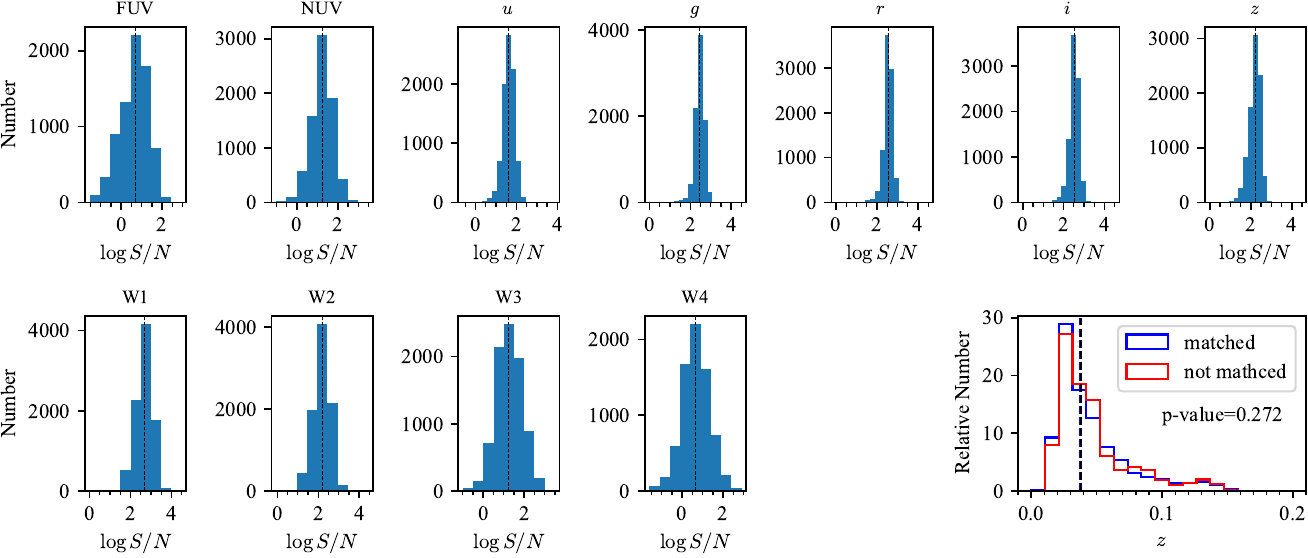} 
 \end{center}
\caption{
Summary of the photometry of the main parent sample.
Histograms show the $\log S/N$ distribution of our sample in each photometric band.
The first row represents FUV, NUV, $u$, $g$, $r$, $i$, and $z$, respectively, from left to right.
The second row represents W1, W2, W3, and W4, respectively, from left to right.
The median $S/N$ value is shown at the top of each panel.
The bottom right panel shows the redshift distribution.
The blue line shows the distribution of MaNGA galaxies matched with the unWISE catalog, i.e., our parent sample, while the red line shows the distribution of MaNGA galaxies not matched with the unWISE catalog. 
The p-value from the Kolmogorov–Smirnov (KS) test is shown in the panel.
}
\label{fig;sample_summary}
\end{figure*}

For SED fitting, we use both photometric data from UV to MIR and IFU-flattened 1D spectroscopic data (see Section \ref{s24}) to obtain more accurate SFHs than the existing SED fitting results, such as Pipe3D \citep{Sanchez2016a, Sanchez2016b, Sanchez2018} and FIREFLY \citep{Goddard2017, Parikh2018}.
For UV (FUV and NUV) photometry from Galaxy Evolution Explorer (GALEX) and optical ($ugriz$) photometry from the SDSS DR8, we use Sérsic fluxes summarized in the NSA and correct them for Galactic extinction using \citet{Schlegel2018}.
We also use IR photometry (W1, W2, W3, and W4) from the unWISE catalog \citep{Schlafly2019}, which is deeper than the ALLWISE catalog due to using coadd images and a forced photometry method \citep{Lang2016}.
Furthermore, because the forced photometry is based on galaxy positions and profiles from the SDSS, the unWISE photometry is more consistent with the NSA photometry described above.
By crossmatching MaNGA galaxies with the unWISE catalog 
with a maximum separation of 10 arcsec, we select 8857 sources with valid photometry (i.e., without missing values) as the main parent sample.
The $S/N$ distribution for each band and the redshift distribution of the sample are shown in figure~\ref{fig;sample_summary}.
We find that crossmatching with the unWISE catalog does not affect the redshift distribution significantly.

\subsection{1D flattening of datacubes}\label{s24}
To transform 3D IFU data (with two spatial and one spectral dimensions) into a form that can be used for SED fitting, for each galaxy, we sum the spectra of all spatial positions to create a 1D spectrum in the following manner.

\begin{figure*}
 \begin{center}
  \includegraphics[width=16cm]{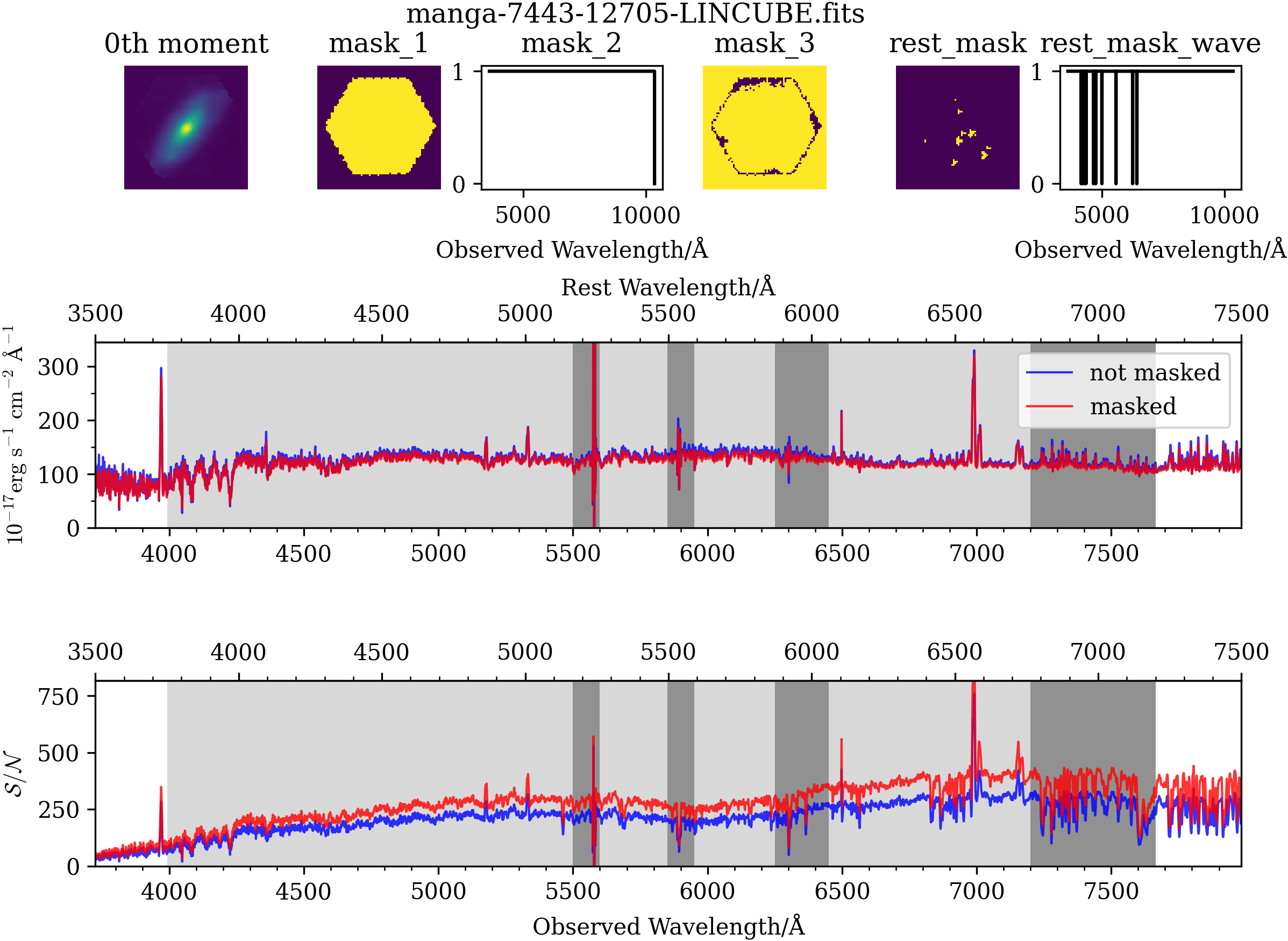} 
 \end{center}
\caption{
An example of our 1D flattening procedure.
The first row shows the 0th moment of an example galaxy and the distributions of masks used in the 1D-flattening procedure.
The second and third rows show a coadded 1D spectrum and its $S/N$ distribution, respectively, with the red (blue) line indicating the result with (without) masking.
}
\label{fig;1dflat}
\end{figure*}

First, based on {\tt MANGA\_DRP3PIXMASK} information, we exclude invalid data with {\tt FORESTAR} or {\tt DONOTUSE} mask.
{\tt FORESTAR} means that the spaxel is contaminated by a foreground star while {\tt DONOTUSE} means that the data of the spaxel are unreliable for some reason.
We do not use a spaxel if it has even one wavelength bin masked with either flag.
The mask\_1 panel in figure~\ref{fig;1dflat} shows the positions removed by this operation in dark blue, and the mask\_2 panel shows the wavelengths removed by this operation (the value of 0 indicates that the wavelength has been removed).

Then, we remove spaxels that either have an invalid value at least one wavelength bin or are located near the edge of the field of view, i.e., at a distance from the center being greater than 0.8 times the maximum distance of all spaxels.
The locations removed by this operation are shown in dark blue in figure~\ref{fig;1dflat}, mask\_3.

In addition, we mask three wavelength ranges [5500~\AA, 5600~\AA], [5850~\AA, 5950~\AA], and [6250~\AA, 6450~\AA] because luminous airglow emission lines contaminate.
These wavelength ranges are shown in dark gray in the spectral and $S/N$ wavelength dependence diagrams in figure~\ref{fig;1dflat}.

Finally, we coadd the spaxels not masked by the above operations to produce 1D-spectrum data.
We calculate the errors in 1D spectra by assuming a normal distribution of errors for each spaxel.
An example coadded 1D-spectrum and its $S/N$ are shown in the second and third lines in figure~\ref{fig;1dflat}.

As mentioned in section~\ref{s21}, the size of the MaNGA field of view varies depending on the sample type.
The field of view covers $1.5R_e$ for the Primary Sample and $2.5R_e$ for the Secondary Sample, where $R_e$ is the effective radius.
However, we do not perform aperture correction according to sample type.
This is because {\sc Prospector} has the function of photospectral correction as explained in section~\ref{s312}.
Note that if there is non-negligible radial dependence in spectra, e.g., strong star formation only outside $1.5R_e$, there may be a bias between the Primary Sample and the Secondary Sample.

\subsection{Value-added catalogs}\label{s25}
The MaNGA sample has various value-added catalogs\footnote{\url{https://www.sdss4.org/dr17/data_access/value-added-catalogs/}} (VAC).
Another advantage of using MaNGA data set is those VACs, and we can discuss the characteristics of RGs in more detail and more easily with such catalogs.

In this study, we use two VACs.
First, we use ``MaNGA PyMorph DR17 photometric catalog'' \citep{Dominguez2022} in section~\ref{s54} to discuss the morphology of the selected RGs.
Second, we use ``MaNGA Pipe3D value-added catalog: Spatially resolved and integrated properties of galaxies for DR17'' (Pipe3D, \cite{Sanchez2022}) in appendix to validate the results of our SED fitting.
We describe the detail of each catalog and the used values in the individual sections.

%%%%%%%%%%%%%%%%% Methodology %%%%%%%%%%%%%%%%%%
\section{Methodology}\label{s3}
\subsection{SED fitting}\label{s31}
SED fitting is a technique of fitting observed data (photometry and/or spectroscopy) with model spectra (for a review, see \cite{Walcher2011}).
Through SED fitting, we can estimate basic parameters of galaxies such as stellar mass ($M^*$), dust attenuation ($\tau$), and star formation history (SFH).

\subsubsection{{\sc Prospector}}\label{s311}
{\sc Prospector} \citep{Leja2017, Johnson2021} is a Baysian-based SED fitting code with which it is possible to fit with a flexible non-parametric SFH (e.g., \cite{CidFernandes2005,Ocvirk2006,Tojeiro2007,Dye2008}) by dividing look-back time into several age bins and assuming a constant SFR in each bin.

With high-quality input data, non-parametric SFH fitting can reproduce the true SFH more accurately than parametric SFH methods because of fewer assumptions on the SFH (e.g., \cite{Leja2019}).
\citet{Lower2020} have suggested that a non-parametric SFH can also estimate $M^*$ more accurately.
However, we should note that non-parametric SFH estimation still has prior dependence as \citet{Leja2019} have demonstrated.
{\sc Prospector} can fit spectroscopic and photometric data together to constrain parameters more tightly than fitting only photometric data \citep{Johnson2021}.

In this study, we input a flattened 1D spectrum and photometric data of each galaxy to {\sc Prospector} and estimate a non-parametric SFH to select RGs.
Below we briefly describe our parameter setting (for a detailed description of {\sc Prospector}, see \cite{Leja2017, Johnson2021}).

\subsubsection{{\sc Prospector} settings}\label{s312}
{\bf \underline{Stellar population:}}
{\sc Prospector} uses the Flexible Stellar Population Synthesis (FSPS, \cite{Conroy2009, Conroy2010}) to generate model SEDs of galaxies.
We assume a \citet{Chabrier2003} initial mass function.
We set stellar metallicity, total stellar mass, and the velocity dispersion of stellar components as free parameters and assume that their priors are uniform in $\log\left(Z^*/Z_\odot\right)$ between $-2$ and 0.19, in $\log\left(M^*/M_\odot\right)$ between $10^8$ and $10^{13}$, and uniform between $10~{\rm km/s}$ and $300~{\rm km/s}$.

{\bf \underline{Nebular emission:}}
We use nebular emission templates in {\sc Prospector} that are based on {\sc CLOUDY} \citep{Ferland1998, Ferland2013} and described in \citet{Byler2017}.
We assume that gas-phase metallicity is equal to stellar metallicity.
We set the ionization parameter $\log U$ as a free parameter with a uniform prior between $-4$ and $-1$.

{\bf \underline{Dust emission:}}
We also use dust emission templates with fixed parameters.
We do not use AGN dust emission templates because we have excluded AGN-hosting galaxies in the sample selection (section~\ref{s22}).

{\bf \underline{Dust attenuation:}}
We assume the two-component dust attenuation model by \citet{Charlot2000}.
This model has two types of attenuation. 
One is the attenuation only for birth clouds, $\tau_{\rm dust,1}\left(\lambda\right)$, calculated as:
\begin{equation}
    \tau_{\rm dust,1}\left(\lambda\right) = \hat{\tau}_{\rm dust,1} \left(\frac{\lambda}{5500~{\rm \AA}}\right)^{-1}.
\end{equation}
The other is the attenuation for the entire galaxy, $\tau_{\rm dust,2}\left(\lambda\right)$.
We use the equation of \citet{Noll2009}: 
\begin{equation}
    \tau_{\rm dust,2}\left(\lambda\right) = \hat{\tau}_{\rm dust,2} \left(k\left(\lambda\right) + D\left(\lambda\right)\right) \left(\frac{\lambda}{5500~{\rm \AA}}\right)^{n},
\end{equation}
where $k\left(\lambda\right)$ is \citet{Calzetti2000} attenuation curve and $D\left(\lambda\right)$ is the Lorentzian-like Drude profile  parameterizing the UV bump.
We use the $D\left(\lambda\right)$ definition by \citet{Kriek2013}.
This attenuation model has three parameters; $\hat{\tau}_{\rm dust,2}$, the ratio of $\hat{\tau}_{\rm dust,1}$ to $\hat{\tau}_{\rm dust,2}$, and the index $n$.
We set all three parameters as free parameters and assume their priors to be uniform between $0$ and $4$ for $\hat{\tau}_{\rm dust,2}$, uniform between $1.0$ and $0.4$ for $n$, and a normal distribution with $\mu=1.0$ and $\sigma=0.3$ truncated at $0.0$ and $2.0$ for $\hat{\tau}_{\rm dust,1}/\hat{\tau}_{\rm dust,2}$.

{\bf \underline{SFH:}}
We fix the number of age bins for non-parametric SFH fitting to eight and calculate the SFR for each bin.
\citet{Leja2019} have examined the dependence of the number of age bins on the reconstructed results and found that there is no significant dependence if the number of age bins is greater than 4.
We also fix the recent five age bins to [$0$ yr, $10^{7.5}$ yr], [$10^{7.5}$ yr, $10^{8}$ yr], [$10^{8}$ yr, $10^{8.33}$ yr], [$10^{8.33}$ yr, $10^{8.67}$ yr], and [$10^{8.67}$ yr, $10^{9}$ yr] in lookback time $t_l$, and set the remaining three age bins by splitting equally $\log t_l$ in the logarithmic interval from $10^9$ yr to $0.95 t_H$.

The typical priors of non-parametric SFHs used in {\sc Prospector} are the continuity prior \citep{Leja2019, Johnson2021} and the Dirichlet prior \citep{Leja2017, Leja2019}.
The continuity prior is for the ratio of SFRs in adjacent age bins, and the Student's-t distribution is widely used.
For the Dirichlet prior, we assume a Dirichlet distribution \citep{Leja2017, Leja2018} for the fractional sSFR in each age bin.
\citet{Leja2019} have examined the prior dependence of SFHs by applying {\sc Prospector} on mock data, finding that both priors can reconstruct the shape of the assumed SFH in mock data. 
However, they have also found that the continuity and Dirichlet priors tend to reconstruct the SFH ``dispersively'' and ``concentratedly'', respectively.
In this study, we use both priors and compare the results.
We have eight age bins, and total stellar mass is also a free parameter.
Thus, the number of free parameters for the SFH is seven for both priors: seven SFR ratios between the adjacent bins for the continuity prior and seven Dirichlet parameters.

{\bf \underline{Spectrophotometric calibration:}}
MaNGA data are already flux-calibrated by the DRP.
However, for any galaxy, the area of IFU spectroscopy is different from photometric apertures.
We apply the spectrophotometric calibration model in {\sc Prospector} to do aperture correction.
This model fits a polynomial calibration function to scale the input observed spectrum to model spectra so that its pseudo photometry is consistent with the observed multiband photometry (\cite{Johnson2021} for detail).
We confirm that this calibration model works well with mock data fitting results.

{\bf \underline{Emission-line marginalization:}}
Because {\sc Prospector} assumes all nebular emission lines to originate from star formation, our SED fitting will fail to reconstruct the spectrum around an observed emission line if the line is contaminated by a LIER, an AGN, or shock heating.
To avoid this problem, we marginalize the amplitude of each emission line in each fitting step. 
We fit for the velocity dispersion of gas, $\sigma_{\rm gas}$, assuming a uniform prior between $10~{\rm km/s}$ and $300~{\rm km/s}$.
We also confirm that introducing line marginalization can improve the model spectrum around a complex single-line feature consisting of narrow emission and broad absorption lines.

{\bf \underline{Likelihood calculation and the outlier model:}}
Finally, we calculate the likelihood $\mathcal{L}$ as the product of the likelihood for the photometric data $\mathcal{L}_{\rm phot}$ and the spectrum $\mathcal{L}_{\rm spec}$, as
\begin{equation}
    \ln\mathcal{L} = \ln\mathcal{L}_{\rm phot} + \left(1-f_{\rm out}\right)\ln\mathcal{L}_{\rm spec} + f_{\rm out}\ln\mathcal{L}_{\rm out},\label{eq_ltot}
\end{equation}
where $\mathcal{L}_{\rm out}$ is the penalty from the outlier model described below (equation \ref{eq_lout}).
Here, $\mathcal{L}$ is calculated on the assumption of Gaussian uncertainties, as
\begin{equation}
    \ln\mathcal{L}\left(f,m,\sigma\right) = -\sum_i^{N} \frac{\left(f_i-m_i\right)^2}{2\sigma_i^2},\label{eq_l}
\end{equation}
where $f$, $\sigma$, and $m$ are the observed flux density, its uncertainty, and the model flux density, respectively.

We apply the outlier model \citep{Hogg2010, Johnson2021} to ignore outlier data not fitted well by the model spectrum (e.g., cosmic rays, airglow lines, and complex line features).
On the assumption that a certain percentage $f_{\rm out}$ of data points are outliers, this model adds a penalty term in the likelihood calculation.
The penalty term is calculated by multiplying the uncertainty by $s_{\rm out}$ for all data points, as
\begin{equation}
    \ln\mathcal{L}_{\rm out}\left(f,m,\sigma\right) = -\sum_i^{N} \frac{\left(f_i-m_i\right)^2}{2\left(s_{\rm out}\sigma_i\right)^2}.\label{eq_lout}
\end{equation}
With this penalty term, the model becomes less sensitive to the spectral data.
In this study, we applied $s_{\rm out}=50$.
\par

Table~\ref{tab;fp} summarizes the free parameters and their priors.
This parameter setting is determined by referring to the run time, the distribution of each parameter obtained from several test runs, the range of each parameter in the Pipe3D and the MaNGA targeting catalog, and settings in some previous studies with {\sc Prospector} \citep{Tacchella2022}.

\begin{figure*}
 \begin{center}
  \includegraphics[width=15cm]{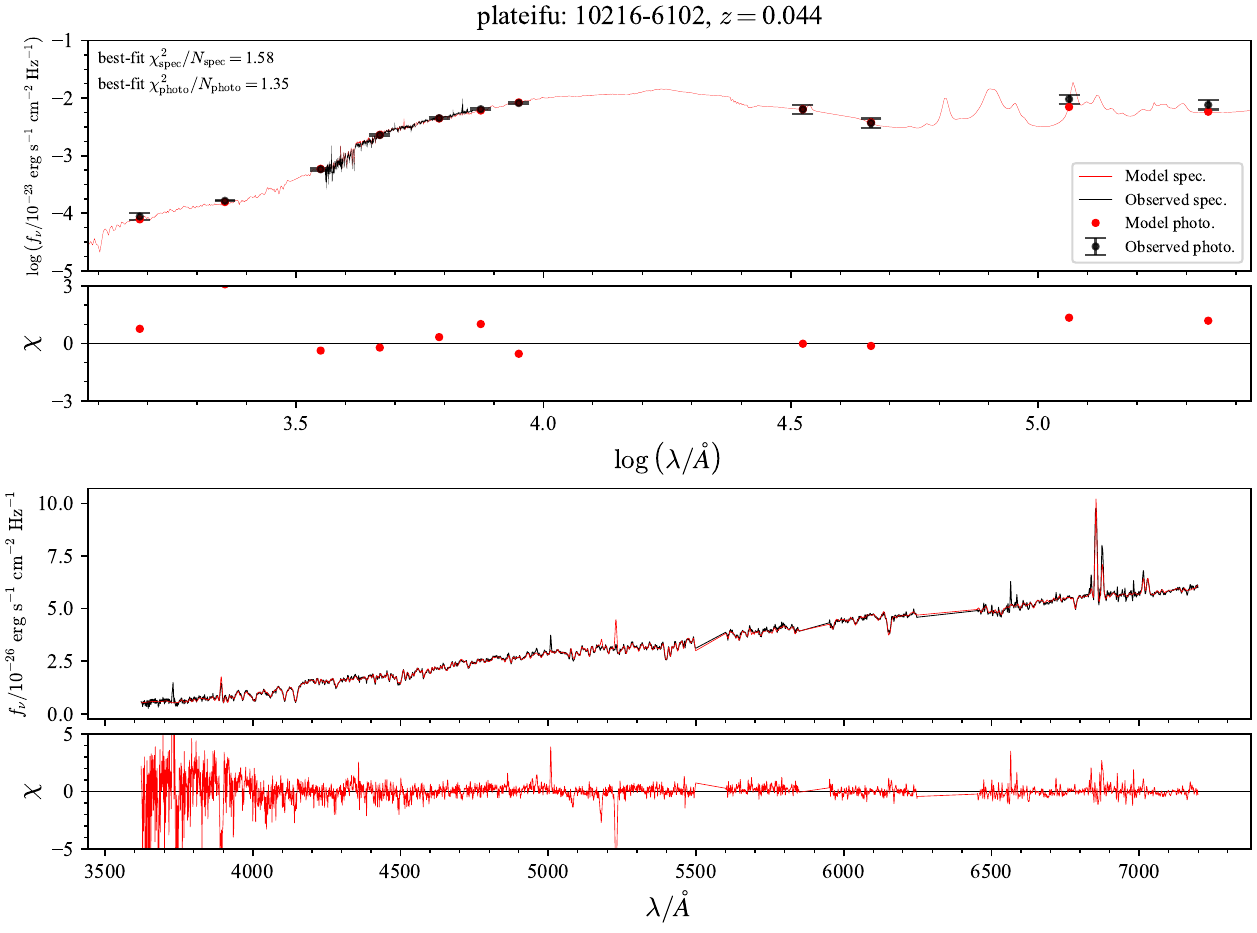} 
 \end{center}
\caption{
The top (bottom) panel compares photometric (spectroscopic) data with the best-fit model for a galaxy at $z=0.044$ (plateifu:10216-6102), with the lower plot in each panel showing $\chi=\left(f_{\nu, {\rm obs}} - f_{\nu, {\rm model}}\right)/\sigma$.}
\label{fig;chi_square}
\end{figure*}

As an example, figure~\ref{fig;chi_square} shows the observed and modeled (best-fit) spectra and photometry of a galaxy at $z = 0.044$ (plateifu:10216-6102).
We calculate $\chi^2$ values for spectroscopic data ($\chi^2_{\rm spec}$) and photometric data ($\chi^2_{\rm photo}$) separately.
We find that the median of $\chi^2/N_{\rm data}$ ($N_{\rm data}$ is the number of data points) for the whole sample is $\chi^2_{\rm spec}/N_{\rm data,spec} = 1.15^{+1.82}_{-0.62}$ and $\chi^2_{\rm photo}/N_{\rm data,photo} = 2.44^{+3.85}_{-1.51}$.
While the median $\chi^2_{\rm spec}/N_{\rm data,spec}$ value is reasonable ($\simeq 1$), we find that objects whose spectra around the short-wavelength end are of poor quality have large $\chi^2_{\rm spec}/N_{\rm data,spec}$.
Indeed, the $\chi^2_{\rm spec}/N_{\rm data,spec}$ of those objects decreases close to unity if the data around the short-wavelength end are excluded.
On the other hand, the large median $\chi^2_{\rm photo}/N_{\rm data,photo}$ value is primarily because the WISE data (especially W4) of most objects are not reproduced well.
This problem may be resolved by adding new free parameters to the dust model.

\begin{figure*}
 \begin{center}
  \includegraphics[width=15cm]{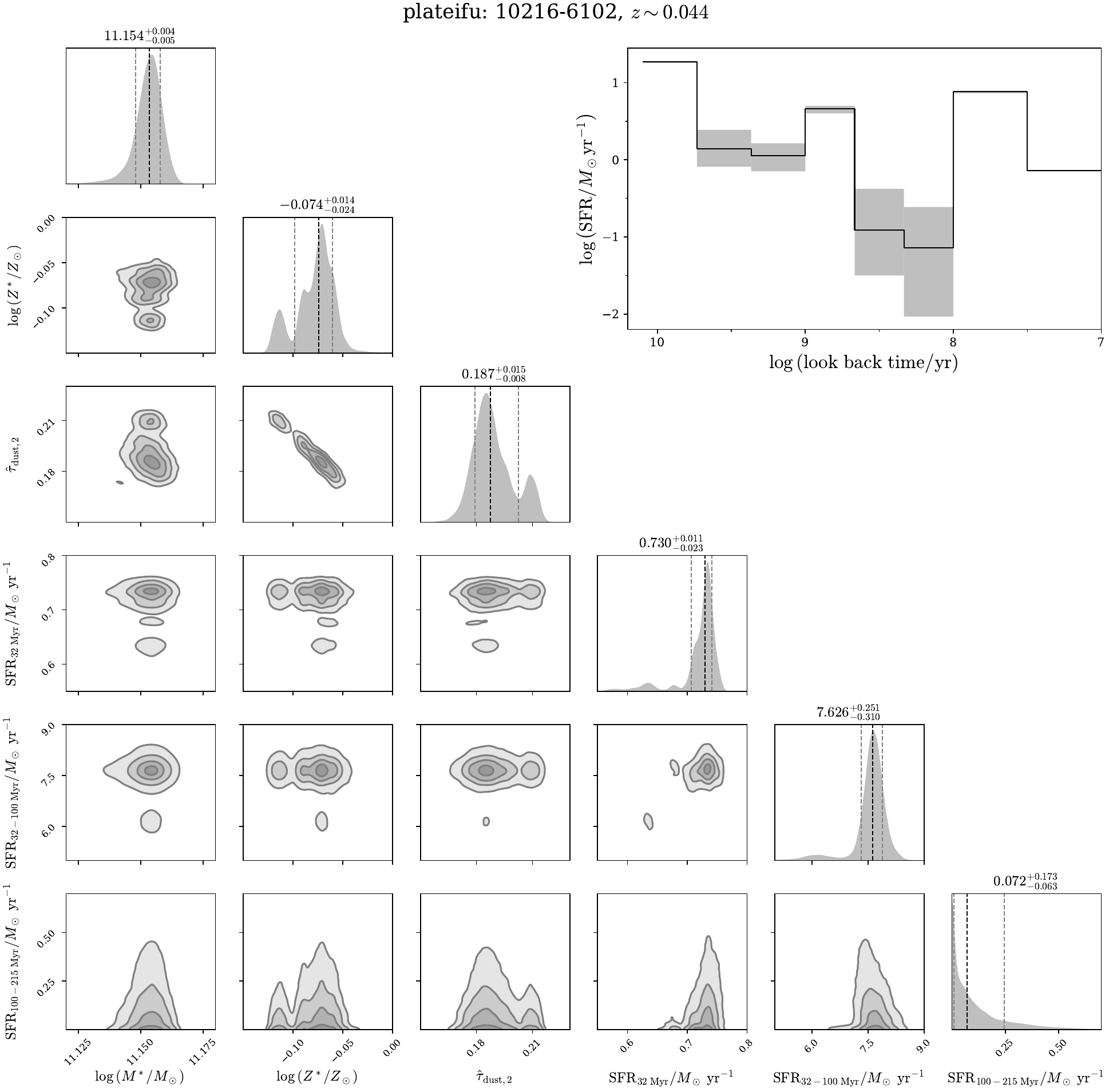} 
 \end{center}
\caption{
Joint posterior distribution plot (corner plot) of $M^*$, $Z^*$, $\hat{\tau}_{\rm dust, 2}$, and SFRs in the recent three age bins ([$0$ yr, $10^{7.5}$ yr], [$10^{7.5}$ yr, $10^{8}$ yr], and [$10^{8}$ yr, $10^{8.33}$ yr]) for the example galaxy in figure~\ref{fig;chi_square} (plateifu:10216-6102, $z = 0.044$).
The diagonal panels show the posterior distribution of each parameter with black and gray dashed lines indicating the median and the $1\sigma$ confidence level, and the panels below each diagonal panel show correlations with the remaining parameters.
The upper right panel shows the estimated SFH of this galaxy; the solid black line indicates the median SFR, with a gray shaded region corresponding to the $1\sigma$ confidence level of the fit.
}
\label{fig;corner}
\end{figure*}

The joint posterior distribution plot (corner plot) of six key parameters for the same example galaxy as figure~\ref{fig;chi_square} is shown in figure~\ref{fig;corner}.
This galaxy is one of the rejuvenation galaxies selected with the method described in section~\ref{s32}.
We find that basic parameters such as $M^*$, $Z^*$ and $\hat{\tau}_{\rm dust,2}$ are strongly constrained.
The estimated SFH distribution shows that this galaxy experienced rejuvenation at $t_l = 10^8~{\rm yr}$ with ${\rm SFR}_{\rm 32~Myr} \sim 10 M_\odot/{\rm yr^{-1}}$, and the resumed star formation ended in the most recent time bin.
As the strong negative correlation between $\hat{\tau}_{\rm dust,2}$ and $\log\left(Z^*/Z_\odot\right)$ suggests, dust-metallicity degeneracy can also be seen in our results.
However, the $1\sigma$ range of $\hat{\tau}_{\rm dust,2}$ and $\log\left(Z^*/Z_\odot\right)$ is much smaller than the median value of each parameter, suggesting that our method can constrain these parameters reasonably well.
Besides, $\hat{\tau}_{\rm dust,2}$ and $\log\left(Z^*/Z_\odot\right)$ are not strongly correlated with the SFRs in the recent three bins that are very sensitive to the selection of RGs (see section~\ref{s32} for our selection method).
These results support our RG selection method.

We also check whether {\sc Prospector} can select true RGs under this setting by running it on mock data (section~\ref{s4}) and comparing 
the results for the real data with
existing catalogs (appendix).

\begin{table*}[]
\caption{Free parameters and their priors for {\sc Prospector} fitting}\label{tab;fp}
\begin{tabular}{lp{6cm}p{6cm}}
\hline\hline
parameter                 & Description & prior \\ \hline
$\log\left(M^*/M_\odot\right)$ & total stellar mass & Uniform: min=8, max=13 \\
$\log\left(Z^*/Z_\odot\right)$ & stellar metallicity & Uniform: min=-2, max=0.19 \\
$\log U$ & ionization parameter & Uniform: min=-4, max=-1 \\
$n$ & power-law index of the dust attenuation curve of the diﬀuse dust & Uniform: min=-1, max=0.4 \\
$\hat{\tau}_{\rm dust,2}$ & optical depth of the diffuse dust attenuation & Uniform: min=0, max=4 \\
$\hat{\tau}_{\rm dust,1}$ & optical depth of the birth-cloud dust attenuation & Clipped normal in $\hat{\tau}_{\rm dust,1}/\hat{\tau}_{\rm dust,2}$: min=0, max=2, $\mu=1$, $\sigma=0.3$\\
$\sigma_*$ & velocity dispersion of the stellar component & Uniform: min=$10~{\rm km/s}$, max=$300~{\rm km/s}$ \\
$\sigma_{\rm gas}$ & velocity dispersion of gas & Uniform: min=$10~{\rm km/s}$, max=$300~{\rm km/s}$ \\
$f_{\rm out}$ & outlier fraction of spectral data points & Uniform: min=$10^{-5}$, max=$0.5$ \\
SFR ratios & $\Delta\log {\rm SFR}$ between adjacent age bins (only when using the continuity prior) & Student’s-t distribution with $\sigma=0.3$ and $\nu=2$ \\
Dirichlet parameters & weight of SFRs in each age bin (only when using the Dirichlet prior) & Dirichlet distribution with $\alpha_D=1$\\
\hline
\end{tabular}
\end{table*}

\subsection{Selection method}\label{s32}
As introduced in section~\ref{s1}, RGs are galaxies that have restarted star formation after being quenched.
We select RGs using reconstructed SFHs.

\begin{figure}
 \begin{center}
  \includegraphics[width=8cm]{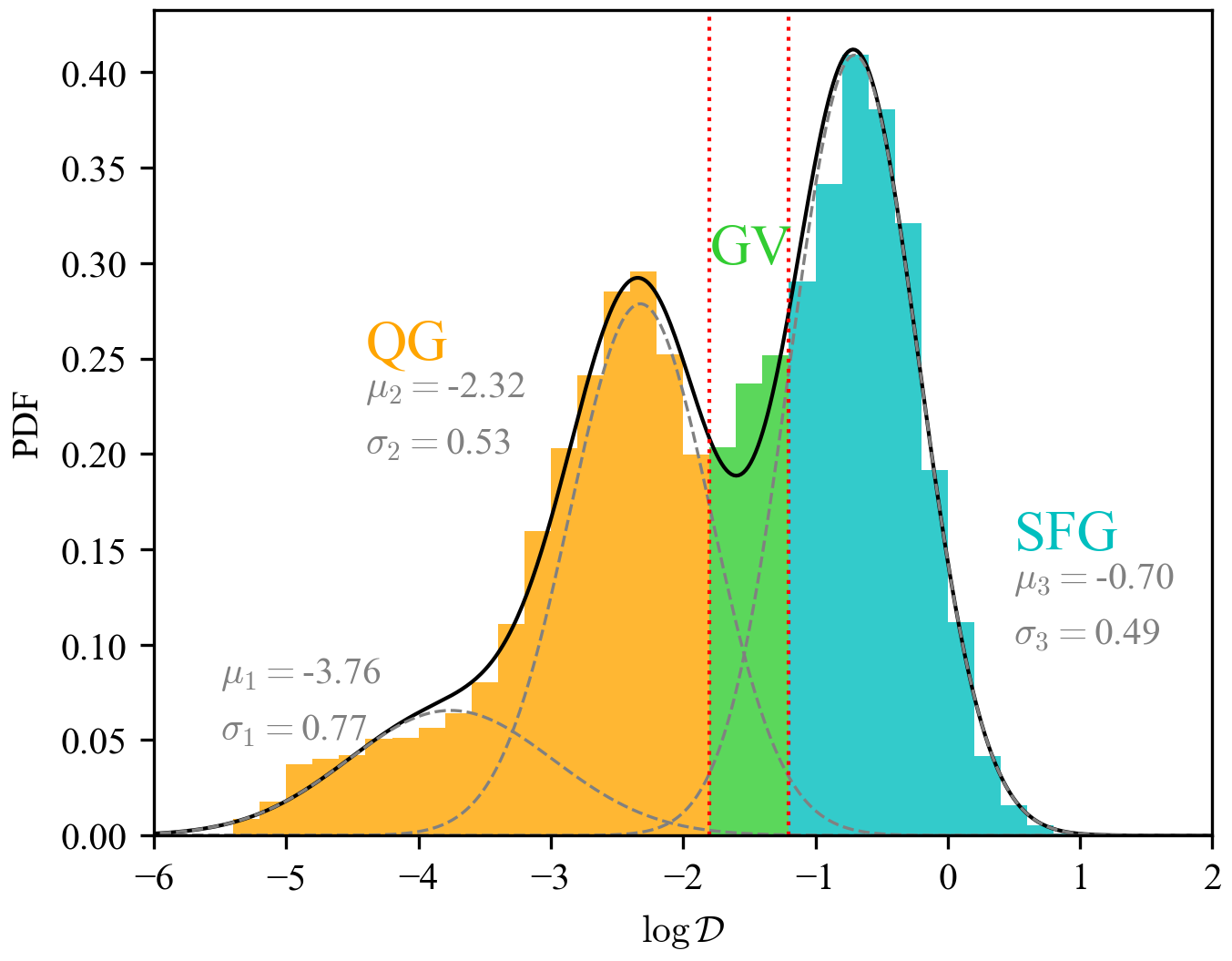} 
 \end{center}
\caption{
The colored histogram is the distribution of $\mathcal{D}$ values in the most recent age-bin obtained with the continuity prior.
The solid black line indicates the result of mixture Gaussian fitting to the histogram, with three gray dashed lines being the three Gaussian components.
The best-fit values of each Gaussian are written in gray.
Red dotted lines indicate the threshold $\mathcal{D}$ values and the histogram is colored according to the classification (orange, green, and cyan for QGs, GVs, and SFGs).
}
\label{fig;threshold}
\end{figure}

For each galaxy, first, we calculate $\mathcal{D}\left(z\right)$, the fraction of the mass that will be formed by the given redshift $z$ if the star formation rate is constant with ${\rm SFR}\left(z\right)$, as
\begin{align}
    \mathcal{D}\left(z\right) &= {\rm sSFR}\left(z\right) \times t_H \left(z\right)\\
    &= \frac{{\rm SFR}\left(z\right)}{M^*\left(z\right)} \times t_H \left(z\right), \nonumber
\end{align}
where $t_H(z)$ is the age of the universe at $z$.
Then, we classify the galaxy in each age bin into an SFG, GV, or QG using $\mathcal{D}\left(z\right)$ to track the evolution of type with cosmic time.
We determine the threshold $\mathcal{D}$ for classification by fitting the $\log\mathcal{D}$ distribution with a three-component mixture Gaussian profile.
Figure~\ref{fig;threshold} shows the fitting result of the $\log \mathcal{D}$ distribution in the latest age-bin, i.e., the bin from $t=10^{7.5}\ {\rm yr}$ to the observed time for the continuum prior. 
We have also confirmed that the Dirichlet prior gives almost the same results.
We fix the SFG/GV and GV/QG boundaries at $\log\mathcal{D}=-1.2$ and $-1.8$, respectively, based on the $1\sigma$ values of the peaks corresponding to SFGs and QGs.

Because the location of the SFMS depends on $z$ (e.g., \cite{Noeske2007, Elbaz2007, Whitaker2012, Speagle2014, Renzini2015, Tomczak2016, Pearson2018, Popesso2023}), the threshold $\mathcal{D}$ value can vary with redshift.
For example, \citet{Tacchella2022} have used $1/3$ and $1/20$ for the $\mathcal{D}$ thresholds ($-0.48$ and $-1.3$ for $\log\mathcal{D}$ thresholds) with a $z\sim0.8$ sample.
However, because our RG selection only focuses on the most recent $10^8$ years much shorter than the timescale of the change of the SFMS location, we can ignore the redshift dependence of the $\mathcal{D}$ threshold value in this study.

Finally, we define an RG as a galaxy that returned to a GV or SFG from a QG within the recent $1 \times 10^8~{\rm yr}$.
This means $t_{\rm rej}$, the lookback time when a rejuvenation started, is either $10^8~{\rm yr}$ or $10^{7.5}~{\rm yr}$ in this study.
In other words, we define an RG as follows,
\begin{itemize}
    \item $t_{\rm rej}=10^{7.5}~{\rm yr}$\\
    $\left[0, 10^{7.5}~{\rm yr}\right]$: GV or SFG\\
    $\left[10^{7.5}~{\rm yr}, 10^{8.0}~{\rm yr}\right]$: QG
    \item $t_{\rm rej}=10^{8.0}~{\rm yr}$\\
    $\left[0, 10^{7.5}~{\rm yr}\right]$: QG, GV or SFG\\
    $\left[10^{7.5}~{\rm yr}, 10^{8.0}~{\rm yr}\right]$: GV or SFG\\
    $\left[10^{8.0}~{\rm yr}, 10^{8.33}~{\rm yr}\right]$: QG
\end{itemize}
It is conceivable that galaxies restart star formation before $10^8~{\rm yr}$.
However, the age-bin setting in our {\sc Prospector} run is equally spaced in logarithmic space; thus, returning to a GV or SFG in an age-bin older than $10^8{\rm yr}$ is a different-timescale event from the above definition.
For this reason, we limit $t_{\rm rej}$ to $10^{7.5}$ and $10^{8.0}$ and focus on the rejuvenation in the recent $\sim100~{\rm Myr}$.

We apply the above selection method to chain data obtained from {\sc Prospector} runs and calculate $p_{\rm RG}$, the probability that the galaxy is an RG, as
\begin{equation}
    p_{\rm RG} = \frac{n_{\rm chain,RG}}{n_{\rm chain}},
\end{equation}
where $n_{\rm chain}$ is the length of all chains and $n_{\rm chain,RG}$ is the number of chains classified as an RG.
As mentioned above, our RG definition has two patterns in terms of $t_{\rm rej}$, and we calculate $p_{\rm RG}$ as the total RG probability; 
as the sum of the probability of each $t_{\rm rej}$ pattern to select galaxies more likely to have rejuvenated regardless of their $t_{\rm rej}$.
We define the $t_{\rm rej}$ of a given galaxy as the timing with a higher $p_{\rm RG}$.

Each galaxy has two $p_{\rm RG}$ values: one with the continuum prior, $p_{\rm RG,con}$, and the other with the Dirichlet prior, $p_{\rm RG,Dir}$.
We select galaxies with $p_{\rm RG,con}>0.8$ and $p_{\rm RG,Dir}>0.5$ as RGs.
We determine these criteria with the SED fitting results of our parent sample and the mock data as described in sections~\ref{s43}.

\begin{figure*}
 \begin{center}
  \includegraphics[width=17cm]{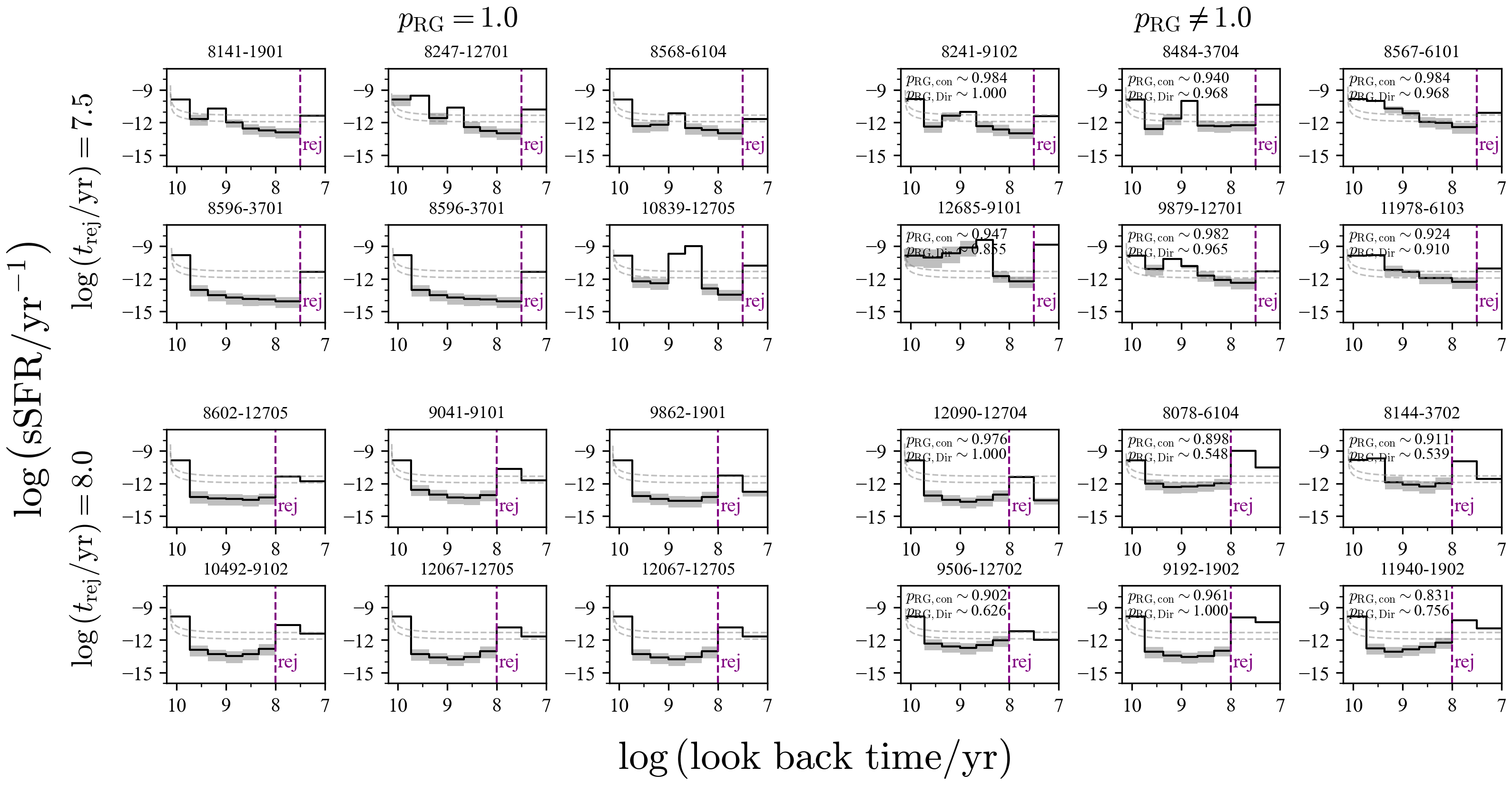} 
 \end{center}
\caption{
Examples of the sSFHs of selected high-$p_{\rm RG}$ RGs 
with $p_{\rm RG}=1$ (rows 1, 2, and 3) 
and $p_{\rm RG} \neq 1$ (rows 4, 5, and 6, where $p_{\rm RG}$ is given in each panel).
Lines 1 and 2 are for objects with $\log\left(t_{\rm rej}/{\rm Myr} \right) = 7.5$ while lines 3 and 4 for $\log\left(t_{\rm rej}/{\rm Myr}\right) = 8$, where $t_{\rm rej}$ is indicated by a purple dashed line.
In each panel, the solid black line indicates the median sSFR, 
with a gray shaded region corresponding to the $1\sigma$ confidence level of the fit.
The gray dashed lines in each panel show the SFG/GV/QG thresholds.
}
\label{fig;sSFH}
\end{figure*}

Figure~\ref{fig;sSFH} shows examples of the specific star-formation history (sSFH) for the selected RGs.
Note that our method depends on the classification into QG/GV/SFG and hence that a galaxy with a clearly increasing sSFR may have a relatively low $p_{\rm RG}$ because its sSFR before rejuvenation is just below the threshold between QGs and GVs (e.g., 8078-6104 in figure~\ref{fig;sSFH} column 5 row 3). 

%%%%%%%%%%%%%%%%% Results-1 %%%%%%%%%%%%%%%%%%
\section{Verification of our RG selection Method}\label{s4}
We run {\sc Prospector} with mock data to verify that our method works well.
Since SED fitting with a non-parametric SFH has been pointed out to have prior dependence \citep{Leja2019}, we also discuss prior dependence.

\subsection{Making mock data}\label{s41}
We make mock data at $z=0.1$ with {\sc Prospector}.
Since we also use {\sc Prospector} in the SED fitting (see section~\ref{s31} for the fitting process), we cannot evaluate the validity of the model spectra in this validation.
The purpose of using a mock sample is only to discuss whether our fitting process can reconstruct the correct SFH and select RGs even with noise.

We construct two mock samples, a single-peak-SFH sample and a double-peak-SFH sample, based on the delayed-tau SFH, one of the popular parametric SFHs.

{\bf \underline{Double-peak-SFH mock sample:}}
To examine if our selection method can select RGs and distinguish RGs from non-rejuvenation galaxies (nRGs) with RG-like SFHs, we make mock data with double-peak SFHs that are a superposition of two delayed-tau models as follows,
\begin{align}
    {\rm SFH}\left(t\right) = C_1 &\left(t_1-t_l\right) \exp\left(-\frac{t_1-t_l}{\tau_1}\right) + \nonumber \\ 
    &
    \begin{cases}
      C_2 \left(t_2-t_l\right) \exp\left(-\frac{t_2-t_l}{\tau_2}\right)  & \left(t_l>t_1\right) \\
      0 & \left(t_l >t_2\right),
    \end{cases}
    \label{eq:sfh}
\end{align}
where $\tau_1$ ($\tau_2$) is the time scale of the first (second) delayed tau model, $t_1$ ($t_2$ [$<t_1$]) is the lookback time when the first (second) delayed tau star formation starts, and $C_1$ ($C_2$) is a coefficient of the first (second) delayed tau model.
We calculate $C_1$ and $C_2$ from $f$, the ratio of the formed mass in the second delayed tau SFH to the total $M^*$.
We fix $t_1$ as $0.95 t_H$, and $M^*$ as $10^{11} M_\odot$.
In equation~(\ref{eq:sfh}), the first term shows the typical galaxy evolution scenario of quenching after active star formation. The second term corresponds to a rejuvenation or pseudo-rejuvenation event.
Note that the first term has a negative value in $t_l>t_1$; thus, we calculate the SFH only for $t_l>t_1$.

To examine various RGs and pseudo RGs, we generate 144 SFHs as a combination of the following parameters,
\begin{itemize}
    \item {\bf $f$}: $10^{-1}$, $10^{-2}$, $10^{-3}$, $10^{-4}$
    \item {\bf $\tau_1$}: 1 Gyr, 1.5 Gyr, 2 Gyr
    \item {\bf $\left(t_H-t_2\right)/\tau_2$}: 1, 10, 50
    \item {\bf $t_H-t_2$}: $10^{7}$ yr, $10^{7.5}$ yr, $10^{8}$ yr, $10^{8.5}$ yr. 
\end{itemize}

\begin{figure*}
 \begin{center}
  \includegraphics[width=16cm]{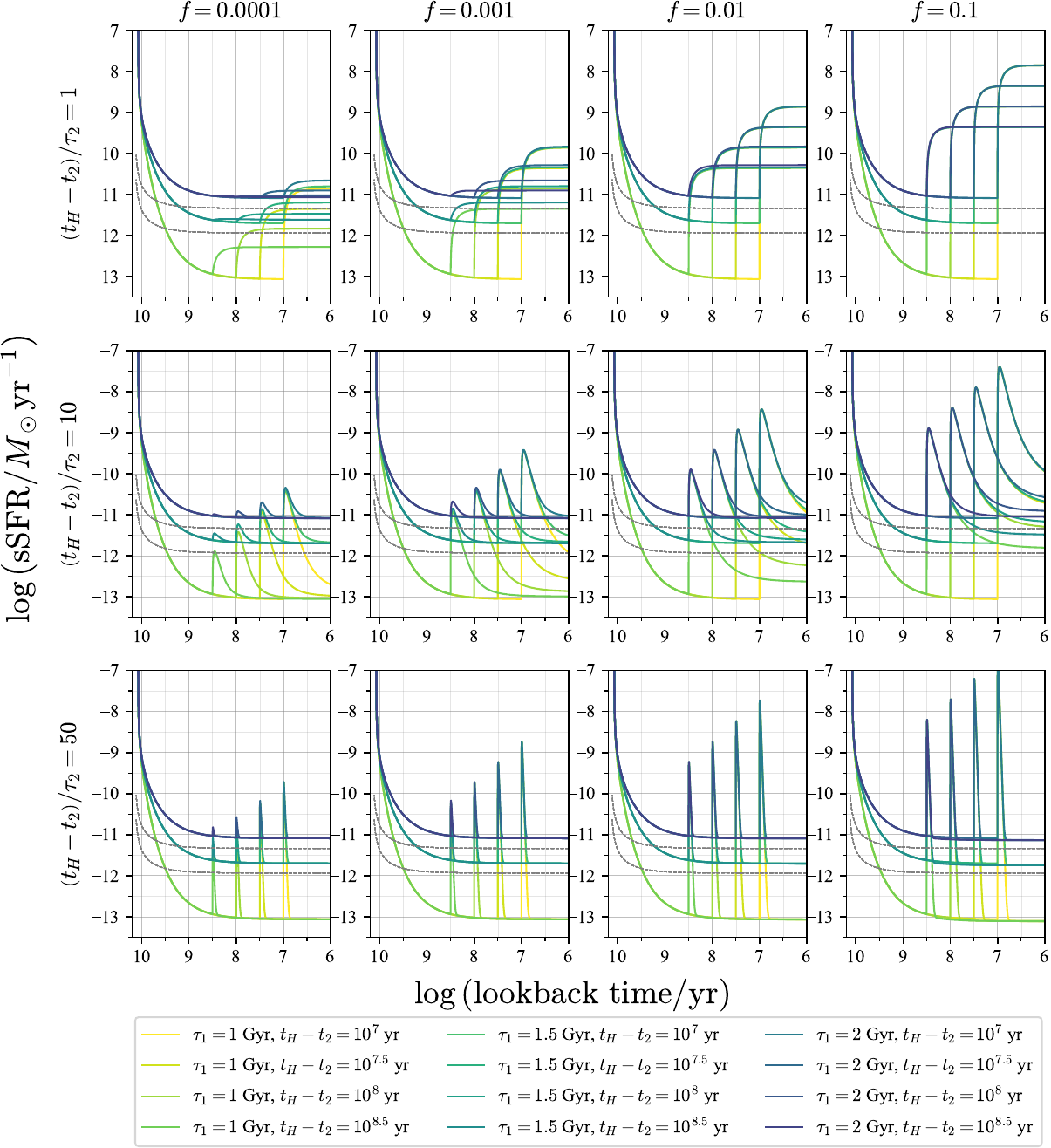}
 \end{center}
\caption{
The sSFHs of the double-peak-SFH mock sample.
Rows 1, 2, 3, and 4 are for $f=0.0001$, $0.001$, $0.01$, and $0.1$, respectively, while lines 1, 2, and 3 for $\left(t_H-t_2\right)/\tau_2=1$, $10$, and $50$, respectively.
Different colors of lines correspond to different combinations of $\tau_1$ and $t_H-t_2$.
Gray dashed lines show the 
boundaries of SFG/GV/QG classification.
}
\label{fig;mock_RG}
\end{figure*}

The specific star-formation histories (sSFHs) of the mock data are shown in figure~\ref{fig;mock_RG}.
Note that although these mock galaxies have two star-formation peaks, not all mock galaxies are RGs.
For example, if we apply our definition introduced in section~\ref{s32}, some galaxies were SFGs or GVs just before the start of the second star formation, and these are not classified as RGs.
Besides, we should also note that some galaxies with $f\sim0.1$ have extremely high SFRs such as $\log\left({\rm SFR}/M_\odot\ {\rm yr^{-1}}\right)\gtrsim3$.

We calculate mock SEDs with the above modeled SFHs.
We add dust and nebular emissions to the stellar continuum.
We fix $\log\left(Z/Z_\odot\right)$ at $-0.5$ and $\hat{\tau}_2$ at $ 0.3$, which are similar to the estimated values for our galaxy sample as shown in appendix.
We assume a \citet{Chabrier2003} initial mass function as in the case of real data.
Then, we use the mock SEDs to calculate mock photometric and spectral data by adding noise following the $S/N$ values shown in the table~\ref{tab:sn}.
These $S/N$ are mainly based on the median $S/N$ for the main sample, but we fix the mock $S/N$ to $20$ for $S/N>20$ and $5$ for $S/N>5$ WISE photometry to account for possible systematic errors, such as aperture correction errors.
Thus, our mock data are more conservative than the real data in terms of $S/N$.

\begin{table*}[]
\caption{Median $S/N$ for the main sample and the assumed $S/N$ in making mock observation data.}
\begin{tabular}{llllllllllllll}
       & \multicolumn{2}{c}{GALEX}                         & \multicolumn{5}{c}{SDSS}                                                                                              & \multicolumn{4}{c}{WISE}                                                                          & \multicolumn{2}{c}{MaNGA}                          \\
       & \multicolumn{1}{c}{FUV} & \multicolumn{1}{c}{NUV} & \multicolumn{1}{c}{$u$} & \multicolumn{1}{c}{$g$} & \multicolumn{1}{c}{$r$} & \multicolumn{1}{c}{$i$} & \multicolumn{1}{c}{$z$} & \multicolumn{1}{c}{W1} & \multicolumn{1}{c}{W2} & \multicolumn{1}{c}{W3} & \multicolumn{1}{c}{W4} & \multicolumn{1}{c}{blue} & \multicolumn{1}{c}{red} \\\hline\hline
Median & 5.1                     & 17.6                    & 42.9                  & 290.1                 & 357.6                 & 341.3                 & 167.6                 & 485.7                  & 155.7                  & 17.8                   & 4.5                    & 23.0                     & 46.6                    \\
Mock   & 5.1                     & 17.6                    & 20                    & 20                    & 20                    & 20                    & 20                    & 5                      & 5                      & 5                      & 4.5                    & 20                       & 20                     
\end{tabular}
\label{tab:sn}
\end{table*}

{\bf \underline{Single-peak-SFH mock sample:}}
To discuss the possibility that galaxies without rejuvenation or a second star-formation event are erroneously selected, we also generate mock data with single peak SFHs using the single delayed-tau model as,
\begin{align}
    {\rm SFH}\left(t\right) = C_1 &\left(t_1-t_l\right) \exp\left(-\frac{t_1-t_l}{\tau_1}\right).
    \label{eq:sfh2}
\end{align}
This time, we also change $t_1$: the lookback time of starting star formation.
We generate 25 SFHs as a combination of the following parameters,
\begin{itemize}
    \item {\bf $\tau_1$}: $2.00\times10^{8}$ yr, $3.56\times10^{8}$ yr, $6.32\times10^{8}$ yr, $1.12\times10^{9}$ yr, $2.00\times10^{9}$ yr
    \item {\bf $t_H-t_1$}: $1.25\times10^{9}$ yr, $2.19\times10^{9}$ yr, $3.84\times10^{9}$ yr, $6.74\times10^{9}$ yr, $1.18\times10^{10}$ yr,
\end{itemize}
where $\tau_1$ is equally sampled in log space between $2\times10^8$ yr and $2\times10^9$ yr, and $t_1-t_l$ is equally sampled in log space between $0.1t_H$ and $0.95t_H$.

\begin{figure*}
 \begin{center}
  \includegraphics[width=16cm]{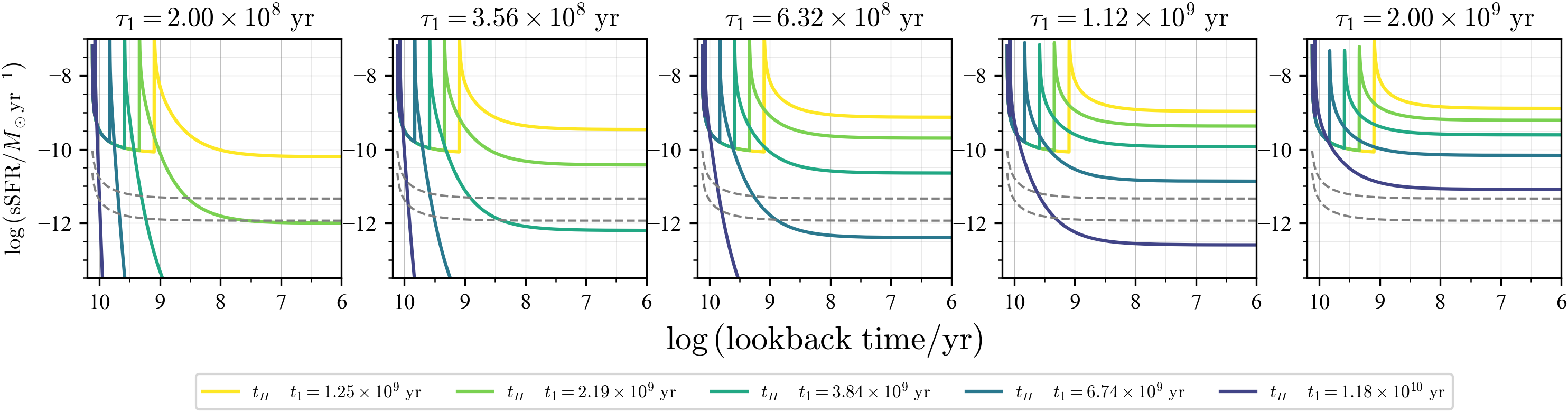} 
 \end{center}
\caption{
The sSFHs of the single-peak-SFH mock sample with $\tau_1=2.00\times10^{8}$ yr, $3.56\times10^{8}$ yr, $6.32\times10^{8}$ yr, $1.12\times10^{9}$ yr, and $2.00\times10^{9}$ yr from left to right.
Different colors of lines indicate different $t_H-t_1$.
Gray dashed lines show the boundaries of SFG/GV/QG classification.
}
\label{fig;mock_nRG}
\end{figure*}

The sSFHs of the single-peak-SFH mock sample are shown in figure~\ref{fig;mock_nRG}.
They have a wide range of current sSFRs over the SFR, GV, and QG regimes and have never experienced a second active star formation, including rejuvenation.

In the calculation of mock SEDs, we make the same assumptions for the metallicity, dust extinction, and IMF as for the double-peak-SFH mock sample. 

\subsection{Basic parameters}\label{s42}
Using the settings described in section~\ref{s31}, we run {\sc Prospector} on the mock galaxies and estimate their SFH and other parameters.

\begin{figure*}
 \begin{center}
  \includegraphics[width=16cm]{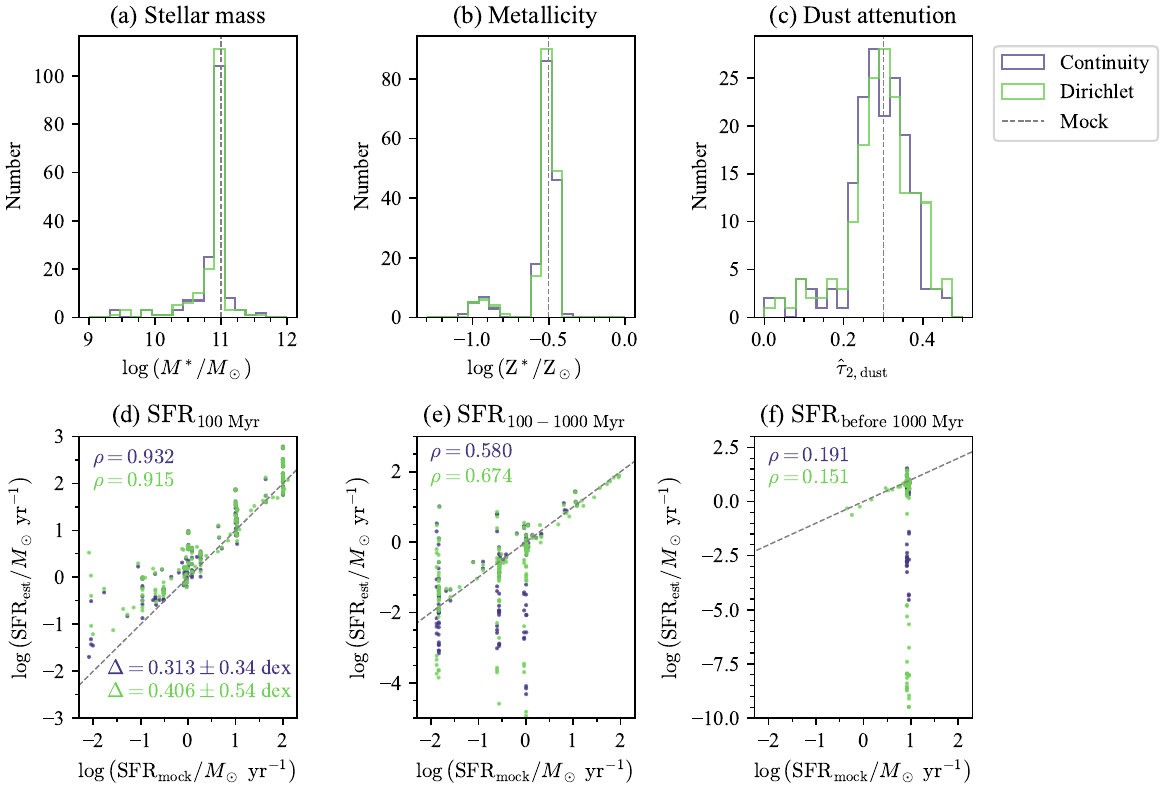} 
 \end{center}
\caption{
Panels (a), (b), and (c) show the histograms of estimated total stellar masses, metallicities, and optical depths of the diffuse dust attenuation, respectively, for the continuity prior in purple and the Dirichlet prior in green.
The assumed values (the true answers) are indicated by gray dashed lines.
Panels (d), (e), and (f) compare estimated SFRs with mock SFRs within 100 Myr, between 100 Myr and 1000 Myr, and before 1000 Myr, respectively, for the continuity prior in purple and the Dirichlet prior in green.
Spearman's correlation coefficient $\rho$ for each prior is shown in the upper left corner of each panel.
The solid black lines show the $y=x$ relation.
We perform fitting with $y=x+\Delta$ only for panel (d).
Generally, Our method reconstructs each parameter well, except for SFR in old age (${\rm SFR_{before~1000~Myr}}$).
}
\label{fig:mock_param}
\end{figure*}

We compare the estimated parameters with the assumed ones in the mock data in figure~\ref{fig:mock_param}.
Figures~\ref{fig:mock_param}~(a), (b), and (c) shows the distribution of $M^*$, ${\rm Z}$, and $\hat{\tau}_{\rm 2, dust}$, respectively.
We find that {\sc Prospector} reproduces the values of these parameters well.
We also find that there are no large differences between the results for the two priors.

Next, we test how well the assumed SFHs are reconstructed by calculating mean SFRs in three bins: $t_l<100~{\rm Myr}$, $100~{\rm Myr}<t_l<1000~{\rm Myr}$, and $t_l>1000~{\rm Myr}$ (figure~\ref{fig:mock_param}~(d), (e), and (f), respectively).
Figure~\ref{fig:mock_param}~(d) compares ${\rm SFR_{100~Myr}}$, the SFR in $t_l<100~{\rm Myr}$, showing a strong correlation between the mock and estimated values.
For galaxies with low assumed SFRs ($\log{\rm SFR_{100~Myr}}\lesssim-0.5$), the continuity prior reproduces the mock values better than the Dirichlet prior.
However, both priors overestimate the SFRs by about 0.3-0.4 dex.

Figure~\ref{fig:mock_param}~(e) compares ${\rm SFR_{100-1000~Myr}}$, the SFR in $100~{\rm Myr}<t_l<1000~{\rm Myr}$, finding a similarly strong correlation.
For galaxies with medium assumed SFRs ($\log{\rm SFR_{100-1000~Myr}}\sim0$), the continuity prior often underestimates ${\rm SFR_{100-1000~Myr}}$.
On the other hand, for galaxies with low assumed SFRs ($\log{\rm SFR_{100-1000~Myr}}\sim-2$), the Dirichlet prior often overestimates ${\rm SFR_{100-1000~Myr}}$.

Figure~\ref{fig:mock_param}~(f) compares ${\rm SFR_{before~1000~Myr}}$, the SFR in $t_l>1000~{\rm Myr}$, and finds a weak correlation in both priors. 
This study does not focus on past star formations.
However, because we define RGs with sSFHs, underestimation of the old stellar population may cause  an overestimation of the current sSFR and affect our selection.
We discuss this problem in section \ref{s43}.

The distribution of estimated metallicities (figure~\ref{fig:mock_param}~(b)) shows a small peak with lower metallicities ($\log\left({\rm Z^*}/{\rm Z_\odot}\right)\lesssim-0.8$) in addition to the primary peak around the assumed value ($\log\left({\rm Z^*}/{\rm Z_\odot}\right)\sim-0.5$).
The galaxies around the small peak have underestimated masses ($\log\left(M^*/M_\odot\right)\lesssim10.7$).
The double-peak-SFH mock sample have 16 galaxies with $\log\left({\rm Z^*}/{\rm Z_\odot}\right)<-0.8$, and 15 of them ($\sim94\%$) have $t_H-t_2=10^7~{\rm yr}$.
Besides, ten and six (about 63 and 38\%) have $f=0.1$ and $0.01$, respectively.
The ${\rm SFR_{before~1000~Myr}}$ of these galaxies are greatly underestimated.
It is likely that {\sc Prospector} cannot reconstruct old star formation due to an extremely large contribution from recent star formation, i.e., the young stellar population.
Thus, {\sc Prospector} may underestimate the stellar mass and metallicity of galaxies with very recent and active resumed star formation.
However, since many $f=0.1$ SFHs have extremely high SFRs over $10^3M_\odot~{\rm yr}^{-1}$ for low-$z$ galaxies, this underestimation is not a serious problem in applying to real data.
However, this underestimation may need to be considered to search for RGs in the high-z universe 
because they have low stellar masses and high SFRs.

\begin{figure*}
 \begin{center}
  \includegraphics[width=15cm]{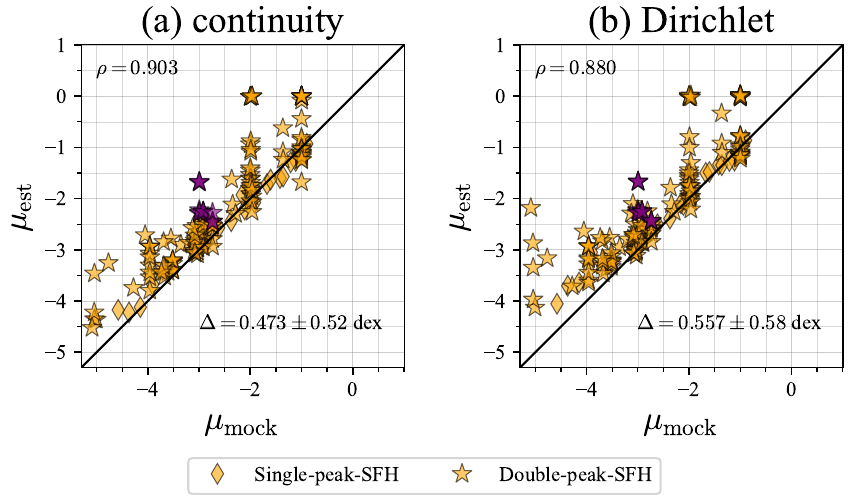} 
 \end{center}
\caption{
Comparison of $\mu_{\rm mock}$ calculated from mock SFHs with $\mu_{\rm est}$ obtained from reconstructed SFHs with the continuity prior (panel [a]) and the Dirichlet prior ([b]).
Diamonds are for single-peak SFH galaxies, and stars for double-peak-SFH galaxies.
Yellow and purple colors mean nRGs and RGs, respectively.
Spearman's correlation coefficient $\rho$ is given in the upper left corner of each panel.
The solid black lines show the $y=x$ relation.
We perform fitting with $y=x+\Delta$, and the value of $\Delta$ and its 68\% error is given in each panel.
Our method can select only $\mu \sim 10^{-3}$ double-peak-SFH objects.
}
\label{fig:mock_mu}
\end{figure*}

To evaluate how well {\sc Prospector} can detect rejuvenation events, we calculate $\mu$, the ratio of the formed mass in the recent 100~Myr to the total stellar mass.
Note that $\mu$ is different from $f$ because we calculate $\mu$ by dividing the formed mass in the recent 100~Myr by the total $M^*$, while calculating $f$ by dividing the formed mass in the second delayed-tau model by the total $M^*$.
Although estimating $f$ from reconstructed SFHs is challenging, $\mu$ can be easily calculated.
Figure~\ref{fig:mock_mu} compares estimated $\mu$ with assumed ones.
We find that {\sc Prospector} tends to overestimate $\mu$ about 0.5~dex.
This tendency is mainly 
due to the overestimation of ${\rm SFR_{100~Myr}}$ (about 0.4~dex higher).

\subsection{RG selection}\label{s43}
\begin{figure*}
 \begin{center}
  \includegraphics[width=17cm]{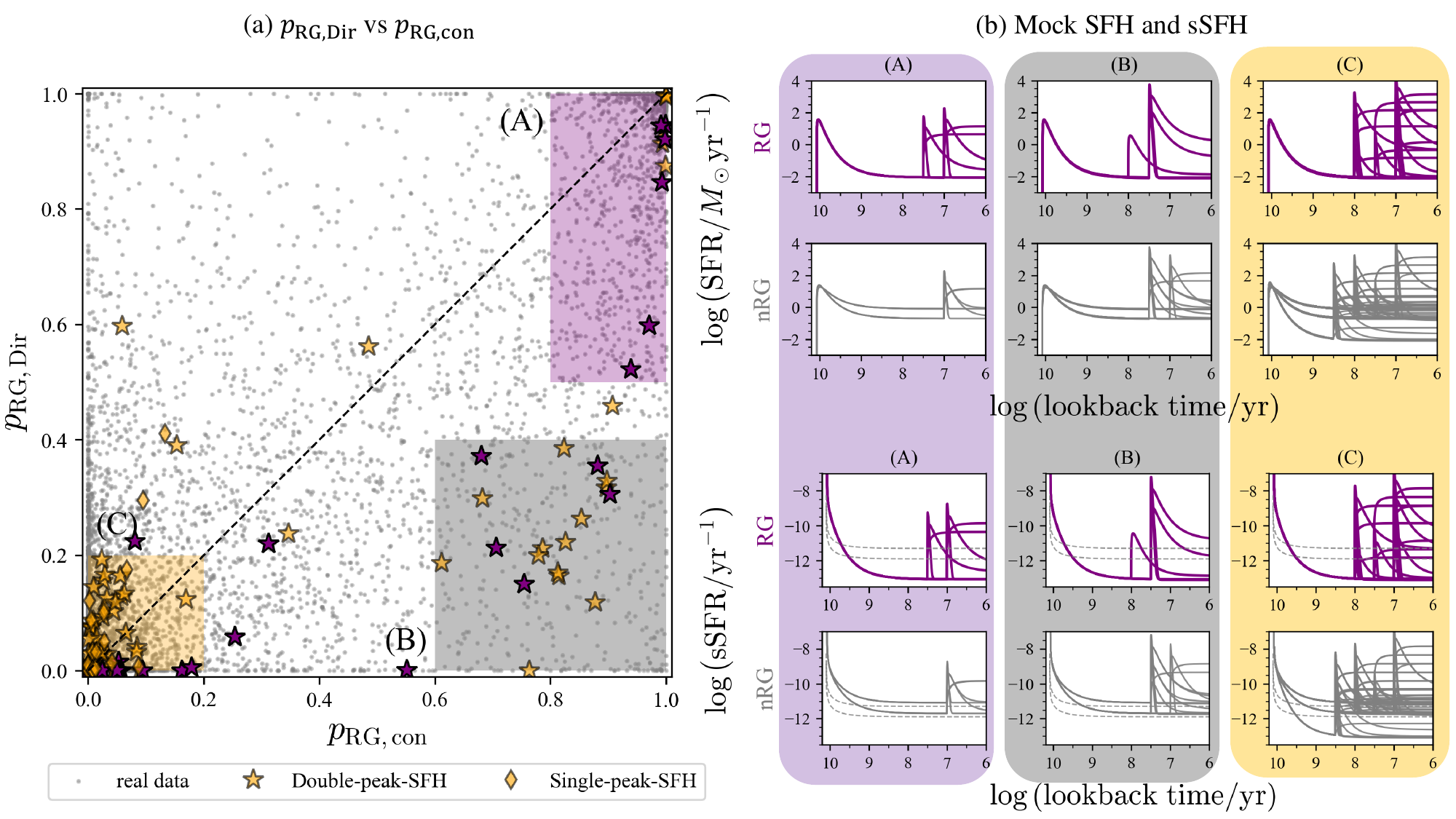} 
 \end{center}
\caption{
(a) Comparison of $p_{\rm RG, con}$ and $p_{\rm RG, Dir}$.
Diamonds, stars, and dots indicate single-peak-SFH mock galaxies, double-peak-SFH mock galaxies, and real data.
Yellow and purple colors mean nRGs and RGs, respectively.
The dashed black lines show the $y=x$ relation.
Regions (A), (B), and (C) are shaded in purple, gray, and yellow, respectively.
Region (A) can select $f=0.001$ RGs with high completeness.
(b) The mock SFHs and sSFHs of mock galaxies in each region. 
RGs and nRGs are separately plotted.
In the sSFH plots, the gray dashed lines indicate the SFG/GV/QG boundaries.
}
\label{fig:mock_rg}
\end{figure*}

The correlation between $p_{\rm RG, con}$ and $p_{\rm RG, Dir}$ for mock and real galaxies is shown in figure~\ref{fig:mock_rg}~(a).
We classify mock galaxies into RGs and nRGs in the same manner as for real galaxies using reconstructed SFHs.
These mock RGs and nRGs are shown in purple and orange, respectively, in figure~\ref{fig:mock_rg}~(a).
We find that almost all the mock RGs have $p_{\rm RG, con} >p_{\rm RG, Dir}$.

For real galaxies, a high Spearman's correlation coefficient of $\rho\simeq 0.79$ is obtained.
However, the value drops to $\rho\simeq 0.38$ if limited to sources with intermediate $p_{\rm RG}$ ($0.1<p_{\rm RG,con}<0.9$), suggesting that the high coefficient value for all galaxies is caused by very RG-like ($0.9<p_{\rm RG}$) and very non-RG-like ($p_{\rm RG}<0.1$) galaxies.
We note that both priors are relatively consistent with each other in that they can correctly select both very RG-like and very non-RG-like objects.

Despite the high correlation coefficient, real galaxies (and mock galaxies) are not evenly distributed around the equality line; objects with medium $p_{\rm RG, con}$ values tend to be lower than the line.
This distribution may be explained as follows.
As seen in section~\ref{s42}, the continuity prior tends to give lower ${\rm SFR}_{\rm 100-1000~Myr}$ and higher ${\rm SFR}_{\rm before~100~Myr}$ than the Dirichlet prior, especially for low SFR objects.
This means that the continuity prior tends to reconstruct more V-shaped, or more RG-like, SFHs than the Dirichlet prior.
Thus, $p_{\rm RG,con}$ is likely to be higher than $p_{\rm RG,Dir}$ especially for low to medium $p_{\rm RG, con}$ objects, 
which are often located around the boundary of QGs and GVs.

Considering these trends in the mock and main parent samples, we set up three regions (A), (B), and (C) in the $p_{\rm RG, Dir}$ - $p_{\rm RG, con}$ plane as candidate regions for RG selection and discuss which region selects RGs best.
Regions (A), (B), and (C) are defined as, 
\begin{itemize}
    \item (A) :$p_{\rm RG,Dir}>0.5$ and $p_{\rm RG,con}>0.8$,
    \item (B) :$p_{\rm RG,Dir}<0.4$ and $p_{\rm RG,con}>0.6$,
    \item (C) :$p_{\rm RG,Dir}<0.2$ and $p_{\rm RG,con}<0.2$.
\end{itemize}
The SFHs and sSFHs of mock RGs and mock nRGs in each region are shown in figure~\ref{fig:mock_rg}~(b).
The parameters of the single-peak-SFH mock sample in each region are also shown in table~\ref{tab:params_region}.

In region (A), we have 12 mock galaxies, and six of them are classified as RGs with mock SFHs.
All six RGs have $f=0.001$ and $t_H-t_2=10^7~{\rm yr},10^{7.5}~{\rm yr}$.
Similarly, all the mock nRGs in region (A) also have $f=0.001$ and $t_H-t_2=10^7~{\rm yr}$.
From these results, galaxies in region (A) are likely to have experienced a very recent second SF with $f\sim0.001$.

Then, in region (B), we have 18 mock galaxies, and five of them are classified as RGs with mock SFHs.
As far as we can judge from the mock data, the contamination rate in region (B) is 72\%, which is higher than in region (A) (50\%).
Region (B) has more mock galaxies with $f=0.01$ and $t_H-t_2=10^{7.5}~{\rm yr}$ than region (A).

Finally, in region (C), we have 106 mock galaxies, and 21 of them are classified as RGs with mock SFHs.
The contamination rate is 80\%, the highest in the three regions.
Most single-peaked-SFH mock galaxies ($22/25$) are distributed in region (C), meaning that our method can correctly diagnose single-peaked-SFH galaxies as nRGs.
We find that mock RGs with $f=0.0001, 0.01,$ and $0.1$ are concentrated in region (C) while there are no mock RGs with $f=0.001$ there.
Besides, most of the mock RGs with $t_H-t_2\gtrsim10^8~{\rm yr}$ are in region (C).
Although our method focuses on rejuvenation within $10^8~{\rm yr}$, rejuvenation events starting from $t_l=10^8~{\rm yr}$ may not be detected in our method.

Based on these results, we choose the region for RG selection.
If we use region (A) to select RGs, the contamination rate of RG selection in mock data is 50\% ($6/12$).
However, the contaminants, i.e., nRGs in region (A), also have a recent second star formation with $f=0.001$.
Besides, region (A) does not include single-peaked-SFH mock galaxies.
Therefore, region (A) can select RGs and RG-like secondary-star-formation galaxies with $f\sim0.001$. 
Since the parameter distribution of mock galaxies should not be the same as the distribution of real galaxies, we cannot calculate the exact contamination rate.
We can only affirm the possibility of selecting such contaminants.

Next, the completeness of RG selection using region (A) is found to be 17\% ($6/36$).
However, similar to the above discussion, we cannot know the exact completeness with our mock data.
For example, as shown in figure~\ref{fig:mock_rg}(b), the mock SFHs in regions (B) or (C) have extremely high SFRs like ${\rm SFR}\gtrsim10^2M_\odot~{\rm yr^{-1}}$.
In the low-$z$ Universe, galaxies with such extremely high SFRs are extremely rare, and the settings of mock data may not be realistic.
If we focus only on galaxies with $f=0.001$, the contamination is $67\%$ ($6/9$).
The results with mock data suggest that using region (A), we can detect weak rejuvenation events with high completeness, although maybe missing galaxies with strong rejuvenation.
From these results, we decide to use region (A).

As for contamination, region (A) may select galaxies with a second star formation whose progenitors are not QGs but GVs or SFGs.
As for completeness, region (A) can miss galaxies that have experienced strong rejuvenation with $f\gtrsim0.01$ or weak rejuvenation with $f\sim0.0001$.
However, region (A) can select RGs with $f\sim0.001$ with high completeness.

We use the results with the continuity prior in the remaining part of this paper.
This is mainly because the continuity prior has a stronger correlation in ${\rm SFR_{100~Myr}}$ (figure~\ref{fig:mock_param} (d)) and in $\mu$ (figure~\ref{fig:mock_mu}).
Besides, $p_{\rm RG, con}$ tends to be higher for $f=0.001$ mock RGs than $p_{\rm RG, Dir}$, suggesting that the continuity prior is better in reproducing $f=0.001$ rejuvenation events.
Note that our selection method uses both $p_{\rm RG, con}$ and $p_{\rm RG, Dir}$ as introduced in section~\ref{s32} and discussed above.

The low completeness of our selection method comes from difficulties in estimating non-parametric SFHs.
For large-$f$ galaxies ($f\gtrsim0.01$), the light from the young stellar population overwhelms that from the old stellar population, i.e., the outshining problem (c.f., \cite{Maraston2010,Sorba2018,Gimenez-Arteaga2023,Narayanan2023}).
In contrast, for small-$f$ galaxies ($f\sim0.0001$), the light from the young stellar population is buried by that from the old stellar population.
It will be important to reduce this uncertainty as much as possible by improving the method, e.g., improving the SED fitting technique and searching for indices to find RGs easily.
The latter will become especially important in the future as data are increasing rapidly.
Because the completeness of selecting $f=0.001$ mock RGs is high, we focus on $f\sim0.001$ RGs in this study.
We will discuss the effects of this incompleteness in section~\ref{s6}.

Again, note that because the mock data are generated with {\sc Prospector}, the comparison in this section cannot demonstrate the validity of model spectra.
To further demonstrate the validity of our SED fitting results with {\sc Prospector}, we also checked that the basic results, such as $M^*$ and SFR, estimated with {\sc Prospector} are strongly correlated with an existing SED fitting result, the Pipe3D DR17 catalog \citep{Sanchez2022}.
The detail of this comparison is introduced in appendix.

\begin{landscape}
\begin{table}[]
\caption{The number of mock galaxies with each mock parameter in each region.}\label{tab:params_region}
\begin{tabular}{ll | llll | lll | lll | llll}\hline\hline
                                         &                    & \multicolumn{4}{c}{$f$}     & \multicolumn{3}{|c|}{$\tau_1$}                               & \multicolumn{3}{|c|}{$\left(t_H-t_2\right)/\tau_2$} & \multicolumn{4}{|c}{$t_H-t_2$}                            \\
                                         &   & 0.0001 & 0.001   & 0.01   & 0.1    & 1 Gyr     & 1.5 Gyr                  & 2 Gyr                   & 1         & 10       & 50      & $10^{7}$ yr      & $10^{7.5}$ yr    & $10^{8}$ yr       & $10^{8.5}$ yr                   \\\hline
\multicolumn{1}{c}{\multirow{7}{*}{All}} & All                & 36  & 36    & 36     & 36     & 48     & 48                    & 48                    & 48        & 48       & 48      & 36     & 36     & 36      & 36                    \\
\multicolumn{1}{c}{}                     & \multirow{2}{*}{A} &0& 12      & 0      & 0      & 6      & 3                     & 3                     & 4         & 4        & 4       & 9      & 3      & 0       & 0                     \\
\multicolumn{1}{c}{}                     &                    &(0\%)& (33\%)  & (0\%)  & (0\%)  & (13\%) & (6.3\%)               & (6.3\%)               & (8.3\%)    & (8.3\%)   & (8.3\%)  & (25\%) & (8.3\%) & (0\%)   & (0\%)                 \\
\multicolumn{1}{c}{}                     & \multirow{2}{*}{B} &0& 2       & 10     & 6      & 5      & 8                     & 5                     & 2         & 9        & 7       & 3      & 14     & 1       & 0                     \\
\multicolumn{1}{c}{}                     &                    &(0\%)& (5.6\%) & (28\%) & (17\%) & (10\%) & (17\%)                & (10\%)                & (4.2\%)   & (19\%)   & (15\%)  & (8.3\%) & (39\%) & (2.8\%) & (0\%)                 \\
\multicolumn{1}{c}{}                     & \multirow{2}{*}{C} &36& 17      & 23     & 30     & 33     & 33                    & 39                    & 38        & 32       & 35      & 23     & 15      & 32      & 35                    \\
\multicolumn{1}{c}{}                     &                    &(100\%)& (47\%)  & (64\%) & (83\%) & (69\%) & (69\%)                & (81\%)                & (79\%)    & (67\%)   & (73\%)  & (64\%) & (42\%) & (89\%)  & (97\%)                \\\hline
\multirow{7}{*}{RG}                      & All                &9& 9       & 9      & 9      & 36     & 0                     & 0                     & 12         & 12        & 12       & 12      & 12      & 12       & 0                     \\
                                         & \multirow{2}{*}{A} &0& 6       & 0      & 0      & 6      & 0                     & 0                     & 2         & 2        & 2       & 3      & 3      & 0       & 0                     \\
                                         &                    &(0\%)& (67\%)  & (0\%)  & (0\%)  & (17\%) & - & - & (17\%)    & (17\%)   & (17\%)  & (25\%) & (25\%) & (0\%)   & - \\
                                         & \multirow{2}{*}{B} &0& 1       & 2      & 2      & 5      & 0                     & 0                     & 0         & 3        & 2       & 0      & 4      & 1       & 0                     \\
                                         &                    &(0\%)& (13\%)  & (22\%) & (22\%) & (14\%) & - & - & (0\%)     & (25\%)   & (17\%)  & (0\%)  & (33\%) & (8.3\%)  & - \\
                                         & \multirow{2}{*}{C} &9& 0       & 5      & 7      & 21     & 0                     & 0                     & 8         & 6        & 7       & 9      & 4      & 8       & 0                     \\
                                         &                    &(100\%)& (0\%)   & (56\%) & (78\%) & (58\%) & - & - & (67\%)    & (50\%)   & (58\%)  & (75\%) & (33\%) & (67\%)  & -\\\hline
\end{tabular}
\end{table}
\end{landscape}

%%%%%%%%%%%%%%%%% Results-2 %%%%%%%%%%%%%%%%%%
\section{Results}\label{s5}
\subsection{RG fraction}\label{s51}
Our method described in section~\ref{s3} selects 1071 RGs out of the $N=8857$ parent sample; thus, the RG fraction calculated by simply dividing the number of RGs by the number of all galaxies is about 12\%. 
Considering the volume weight $w$ provided in the MaNGA targeting catalog \citep{Wake2017} for the Primary+ sample and the secondary sample ({\tt ESRWEIGHT}), we calculate the intrinsic RG fraction, $f_{\rm RG}$, as,
\begin{equation}
    f_{\rm RG} = \frac{\sum_{i\in{\rm RG}}w_i}{\sum_{i\in{\rm All}}w_i}, \label{eq;f_rg_1}
\end{equation}
and obtain $f_{\rm RG}\simeq 8.1\%$.
Furthermore, when restricted to galaxies with $f\sim0.001$ and considering the completeness discussed in section~\ref {s43}, the true RG fraction is $f_{\rm RG}\sim12\%$.
Note, however, that due to the distribution differences between mock galaxies and real galaxies, it is difficult to estimate the true RG fraction using our results and completeness from the mock data, especially when considering RGs with all $f$ values.

As shown in section~\ref{s22}, we have excluded 520 objects as AGN or merging galaxies from the parent sample.
If these galaxies have higher $f_{\rm RG}$ than our parent sample, the $f_{\rm RG}$, including these galaxies, will also be higher than the above value.

\subsection{Images and spectra}\label{s52}
\begin{figure*}
 \begin{center}
  \includegraphics[width=16cm]{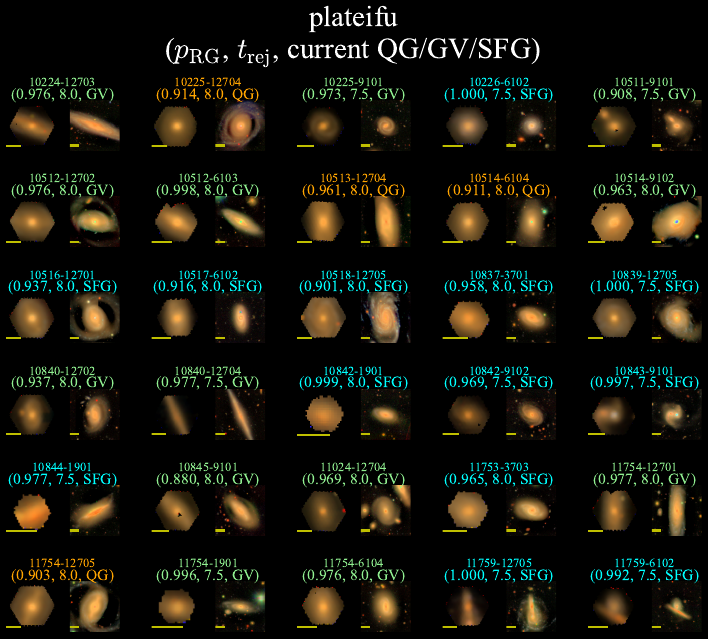} 
 \end{center}
\caption{
Cutout images of 30 RGs included in the HSC-SSP PDR3, with their plateifu given at the top.
The left side of each panel is a $gri$ image constructed from the MaNGA datacube and the right side is a $1' \times 1'$ HSC $gri$ image.
Each image is scaled with the arcsinh method \citep{Lupton2004}.
The yellow line in each panel indicates ten arcsecs.
For each object, $p_{\rm RG}$, $t_{\rm rej}$, and the current state (QG/GV/SFG) are shown below the plateifu, from left to right, with the letters' color corresponding to the current state: orange for QG, yellow-green for GV, and cyan for SFG.
}
\label{fig;cutouts-MaNGAvsHSC}
\end{figure*}

Figure~\ref{fig;cutouts-MaNGAvsHSC} compares images of the selected RGs constructed from MaNGA datacubes with HSC-SSP PDR3 (\cite{Aihara2022}) $gri$ composite images.
HSC images are deeper and cover a larger area than MaNGA datacubes.
We find that although RGs were QGs before rejuvenation, many RGs have a disk structure.
The morphologies of RGs will be analyzed in detail in section~\ref{s54}.

\begin{figure*}
 \begin{center}
  \includegraphics[width=17cm]{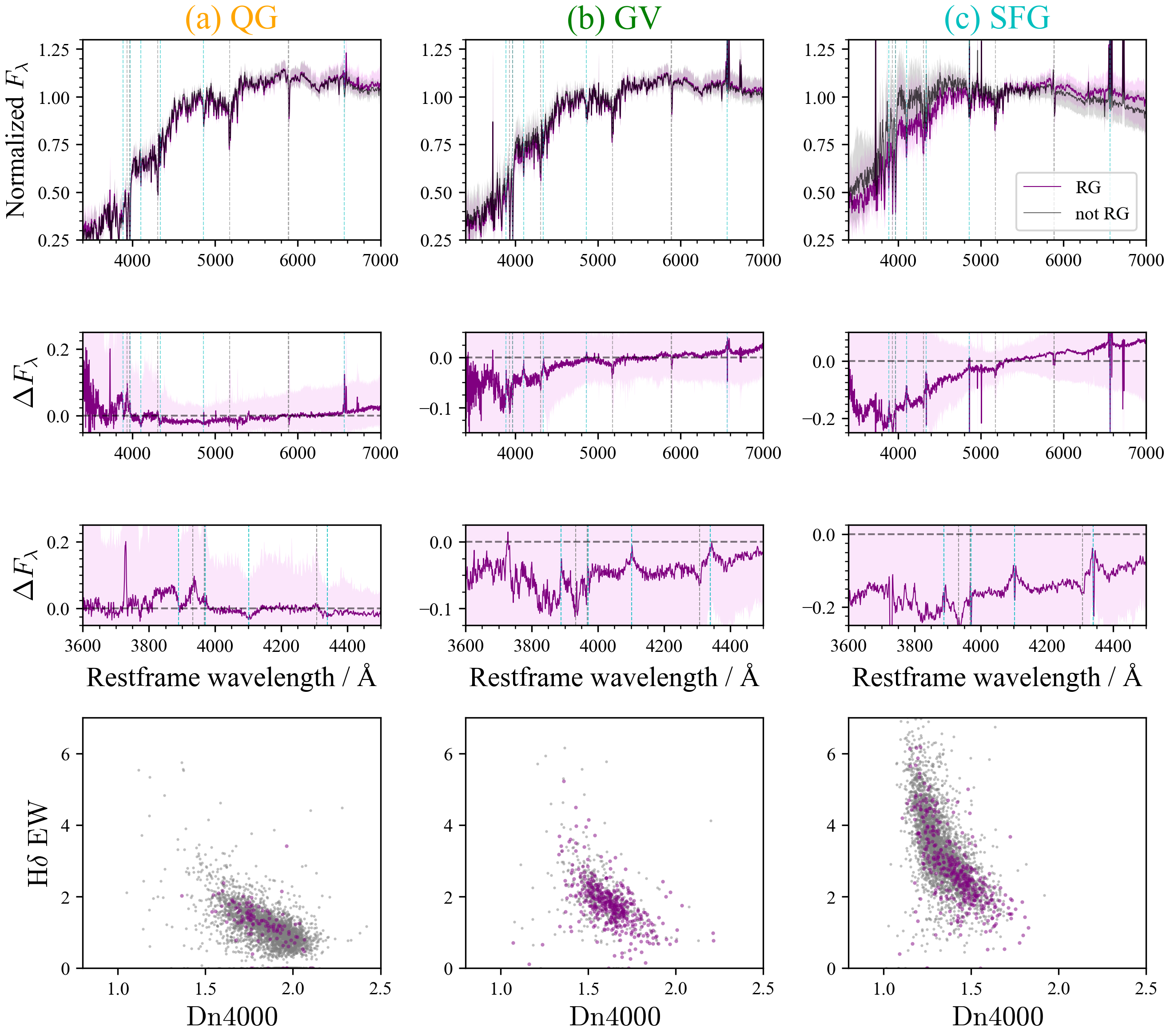} 
 \end{center}
\caption{
(Row 1) Stacked normalized spectra $F_\lambda\left(\lambda\right)$ of QGs (panel [a]), GVs ([b]), and SFGs ([c]).
The purple and black lines show the spectra of those selected and not selected as RGs, respectively, with shaded areas corresponding to the $1\sigma$ confidence intervals.
The wavelengths of several strong Balmer lines and neutral metal lines are indicated with cyan and gray dashed lines, respectively.
The Balmer lines plotted are from H$\alpha$ to H$\zeta$, from left to right.
The metal lines plotted are Na{\sc i}~$\lambda\lambda 5889, 5885$ (D2 and D1 line), Mg{\sc i}~$\lambda 5175$, Ca{\sc i}~$\lambda 4307$ (G line), Ca{\sc ii}$\lambda3968$ (H line), and Ca{\sc ii}$\lambda3933$ (K line). 
(Row 2) Differences in the stacked normalized spectra $\Delta F_\lambda\left(\lambda\right) = \left\{F_{\lambda,{\rm RG}}\left(\lambda\right) - F_{\lambda,{\rm not~RG}}\left(\lambda\right)\right\}/F_{\lambda,{\rm not~RG}}\left(\lambda\right)$ of QGs, GVs, and SFGs (the order is the same with row 1), where the shaded areas correspond to the $1\sigma$ confidence intervals.
Similar to row 1, the wavelengths of Balmer lines and neutral metal lines are indicated.
(Row 3) Magnified views around $4000~{\rm \AA}$.
(Row 4) Relationship between H$\delta$ EW and Dn4000. Gray and purple dots indicate nRGs and RGs, respectively.
}\label{fig;stacked-spec}
\end{figure*}

The upper panels of figure~\ref{fig;stacked-spec} show stacked normalized spectra of current QGs, GVs, and SFGs, separated by whether they are RGs or not; i.e., comparison of rejuvenated QGs (RQGs) with non-rejuvenation QGs (nRQGs), rejuvenating GVs (RGVs) with non-rejuvenation GVs (nRGVs), and rejuvenating SFGs (RSFGs) with non-rejuvenation SFGs (nRSFGs).
Then, to compare the spectra of RGs with those of non-rejuvenation galaxies (nRGs), we calculate a normalized residual spectrum, $\Delta F_\lambda\left(\lambda\right)$, as,
\begin{equation}
    \Delta F_\lambda\left(\lambda\right) = \frac{F_{\lambda,{\rm RG}}\left(\lambda\right) - F_{\lambda,{\rm nRG}}\left(\lambda\right)}{F_{\lambda,{\rm nRG}}\left(\lambda\right)},
\end{equation}
as shown in the lower panels of figure~\ref{fig;stacked-spec}.
Additionally, we examine the distribution of galaxies on the well-studied ${\rm Dn4000}$ versus the ${\rm H\delta}$ equivalent widths (EW) plane in the lowest panels.
We calculate ${\rm Dn4000}$ as follows,
\begin{equation}
    {\rm Dn4000} = \frac{\int_{4000~{\rm \AA}}^{4100~{\rm \AA}} f_\nu~{\rm d}\lambda}{\int_{3850~{\rm \AA}}^{3950~{\rm \AA}} f_\nu~{\rm d}\lambda}.
\end{equation}

Firstly, it is found that the spectral difference between RGs and non-rejuvenation galaxies is less than the $1\sigma$ range, consistent with the previous results from observations by \citet{Chauke2019} and simulated spectra by \citet{Zhang2022}.
As also mentioned in \citet{Zhang2022}, this result suggests that it is difficult to select RGs without an SED fitting method.

Focusing on QGs, panel~(a), $\Delta F_\lambda$ is close to 0 for wavelengths longer than $4000~{\rm \AA}$, but increases to around 0.1 for shorter wavelengths. 
This indicates that RQGs have a slightly weaker $4000~{\rm \AA}$ break than nRQGs.
Similarly, in the ${\rm Dn4000}$ versus ${\rm H\delta}$~EW plane, RQGs have similar EWs to nRQGs but smaller ${\rm Dn4000}$.
This is probably due to the presence of a young stellar population, with a smaller ${\rm Dn4000}$ value, formed in the rejuvenation event.
Also seen is a strong excess in the ${\rm H\alpha}$ emission line, which also indicates recent star formation.

In the case of GVs, panel~(b), $\Delta F_\lambda$ is found to decrease with decreasing wavelength.
This is because RGVs retain a larger population of old stars formed before quenching than nRGVs.
RGVs have a slightly stronger $4000~{\rm \AA}$ break than nRGVs, but not as pronounced as in QGs.
This tendency is also seen in the ${\rm Dn4000}$ versus ${\rm H\delta}$~EW plane, i.e., there is no significant difference in the distributions of RGVs and nRGVs.
Looking at $\Delta F_\lambda$ for each line, H recombination lines (Balmer lines) have larger $\Delta F_\lambda$ than the continuum.
This is likely because RGVs have stronger Balmer emission lines due to active recent star formation.
In contrast, neutral metal lines like Na{\sc i} $\lambda\lambda5885, 5889$ (D1, D2 lines) and Mg{\sc i}$\lambda5175$ have lower $\Delta F_\lambda$ than the continuum, probably because of an aging stellar population formed before quenching.

In the case of SFGs, panel~(c), we can see a stronger decreasing trend of $\Delta F_\lambda$ than in GVs.
This is because RSFGs have a larger amount of old stellar population formed before quenching than nRSFGs.
Similar to GVs, RSFGs have stronger Balmer emission lines and absorption lines of neutral metals than nRSFGs, suggesting the presence of recent star formation and an older stellar population before quenching. 

\subsection{$M^*$ and SFR distributions}\label{s53}
\begin{figure*}
 \begin{center}
  \includegraphics[width=15cm]{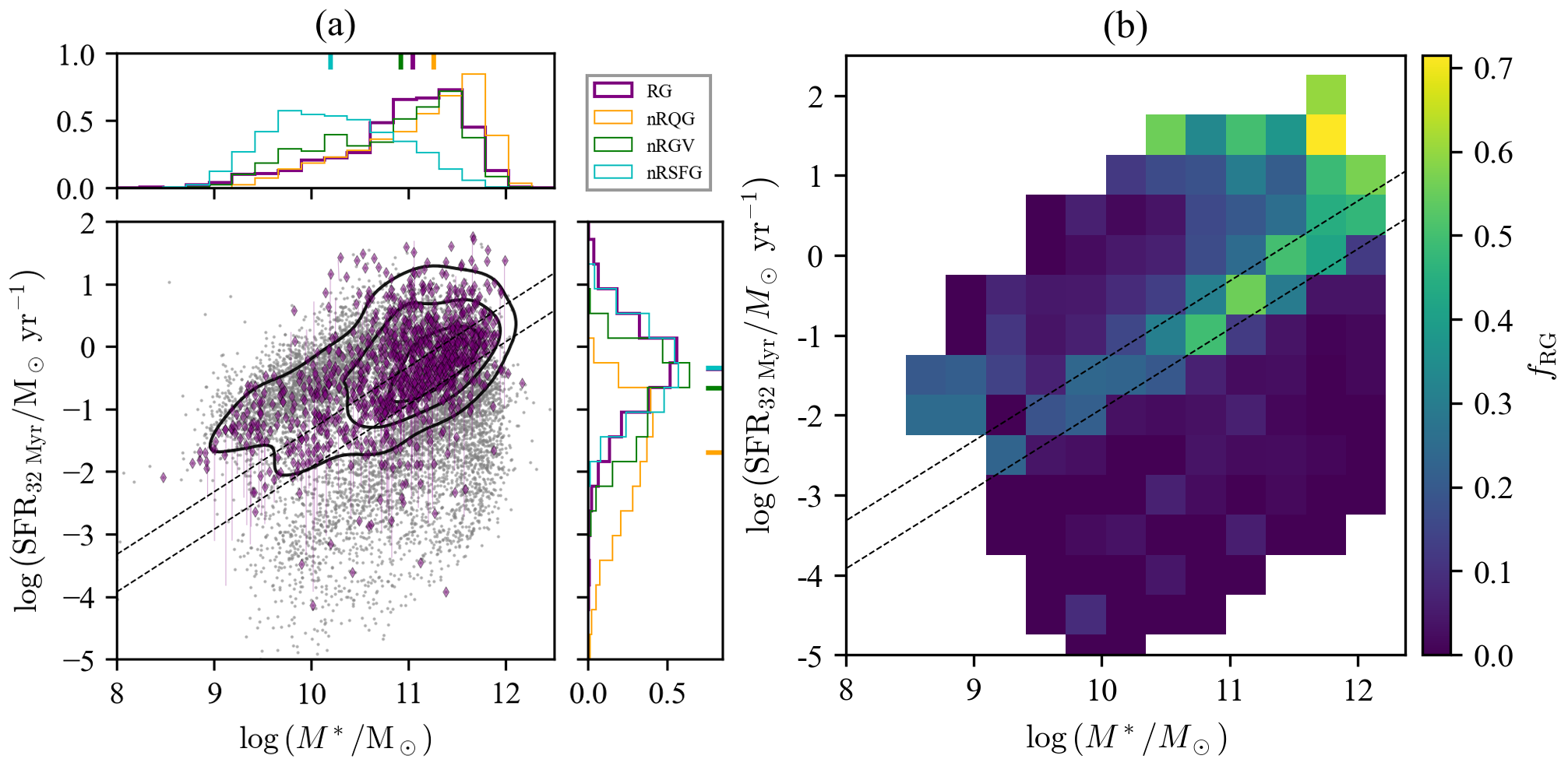} 
 \end{center}
\caption{
(a) The distribution of RGs (purple diamonds) in the SFR-$M^*$ plane, with calculated trajectories during the rejuvenation overplotted as solid purple lines (only for randomly selected 100 RGs with $t_{\rm rej}=10^{7.5}~{\rm yr}$).
The dots show all galaxies in our sample, and black solid contours indicate the 1, 2, and 3 $\sigma$ levels of their distribution.
The two black dashed lines correspond to the boundaries of SFG/GV/QG classification at the median redshift of all galaxies.
The purple, orange, green, and cyan histograms in the subplots are for RGs, not rejuvenation QGs, not rejuvenation GVs, and not rejuvenation SFGs, respectively.
In each subplot, the median of each distribution is indicated by a short bar of the same color as the histogram.
(b) The $f_{\rm RG}$ distribution in the SFR-$M^*$ plane.
The two black dashed lines correspond to the boundaries of SFG/GV/QG classification at the median redshift of the parent sample.
In both panels, we use SFRs in the recent 32~Myr (the most recent age bin).
}
\label{fig;SFRvsM}
\end{figure*}

Figure~\ref{fig;SFRvsM}~(a) shows that the distribution of RGs in the SFR-$M^*$ plane is biased toward high $M^*$ and SFR values, with their median SFR being higher than those of nRSFGs and nRGVs.
We also find that the median $M^*$ of RGs is comparable to that of nRGVs but lower than that of nRQGs.
Note that although RGs have a similar distribution to nRGVs in the 1D distributions of SFR and $M^*$, the distribution of RGs in the SFR-$M^*$ plane is closer to that of SFGs than of nRGVs, as indicated by the black contours in figure~\ref{fig;SFRvsM}~(a).

Figure~\ref{fig;SFRvsM}~(a) also gives an insight into the path in the SFR-$M^*$ plane during rejuvenation events.
Since we focus on recent rejuvenation events within $10^8$~yr (by our definition), even if the sSFR increases to about $10^{-9}~{\rm yr^{-1}}$ due to rejuvenation, the ratio of the formed mass during the rejuvenation to the total stellar mass is only $10^{-1}$.
This means that galaxies move upward almost vertically in the SFR-$M^*$ plane by rejuvenation.
This result is consistent with \citet{Chauke2019}.

Figure~\ref{fig;SFRvsM}~(b) shows the distribution of RG fractions in the SFR versus $M^*$ plane.
The RG fraction is found to be higher for GVs and massive ($M^*>10^{11}M_\odot$) SFGs.
The RG fraction calculated with equation~(\ref{eq;f_rg_1}) for only galaxies with $M^*>10^{11}M_\odot$ is about 15\%, indicating that RGs are more abundant on the high mass side.
These results are consistent with previous studies, which compared rejuvenated galaxies with QGs and found that RGs have a smaller mass than QGs (e.g., \cite{Chauke2019, Tacchella2022}).

\subsection{Morphologies}\label{s54}
\begin{figure*}
 \begin{center}
  \includegraphics[width=15cm]{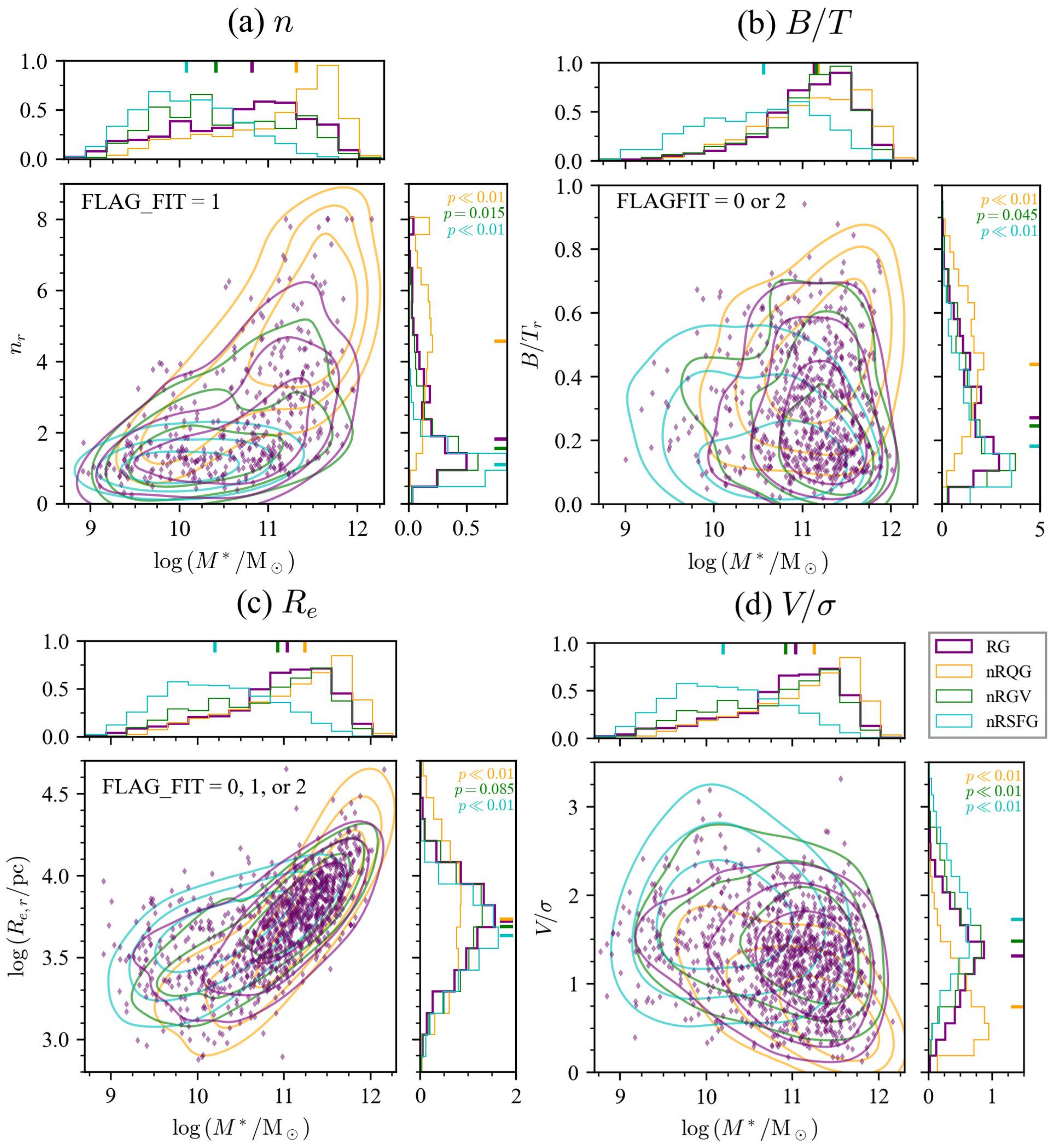} 
 \end{center}
\caption{
The relationship between stellar mass and four indices of morphology: Sérsic index in the $r$ band $n_r$ (panel [a]), bulge to total flux ratio $B/T$ ([b]), effective radius in the $r$ band $r_{e,r}$ ([c]),  the ratio of the rotation velocity $V$ to the velocity dispersion $\sigma$ ([d]).
Plotted in panels (a), (b), and (c) are only galaxies with {\tt FLAG\_FIT}=1, {\tt FLAG\_FIT}=0 or 2, and {\tt FLAG\_FIT}=0, 1, or 2.
In each panel, purple, orange, green, and cyan colors indicate RGs, nRQGs, nRGVs, and nRSFGs, respectively, where dots are individual objects, contours are the 0.5, 1, and 2$\sigma$ levels of the distribution, and histograms are the distributions of individual parameters, with medians indicated by short bars.
Shown in the top right of each $y$-axis histogram are the $p$-values of the Kolmogorov–Smirnov test between the RG sample and the three non-rejuvenation galaxy samples (nQGs, nGVs, nSFGs).
}
\label{fig;morph}
\end{figure*}

In order to compare morphologies between RGs and other galaxies, we show in figure~\ref{fig;morph} the distributions of the $r$-band Sérsic index $n_r$, the $r$-band effective radius $R_{e,r}$, the bulge-to-total flux ratio $B/T$, and the ratio of rotational velocity $V$ to velocity dispersion $\sigma$, $V/\sigma$.
For $n_r$, $R_{e,r}$, and $B/T$, we use the PyMorph catalog \citep{Dominguez2022}, which summarizes photometric morphological parameters obtained from $g$, $r$, and $i$ images of MaNGA DR17 galaxies.
This PyMorph catalog has two types of results, one fit by a Sérsic profile and the other by a Sérsic+exponential profile, with {\tt FLAG\_FIT} parameter indicating which result is reliable: 
{\tt FLAG\_FIT}=0 indicates that both fits are reliable; 
{\tt FLAG\_FIT}=1 indicates that only the Sérsic fit is successful while the Sérsic+exponential profile fit may not be reliable; 
{\tt FLAG\_FIT}=2 indicates that only the Sérsic+exponential fit is successful while the Sérsic profile fit may not be reliable; 
{\tt FLAG\_FIT}=3 indicates both fits may be unreliable.
Using this parameter, we construct a different sample depending on the morphological parameter to be examined.
We use only galaxies with {\tt FLAG\_FIT}=1 when discussing $n_r$, only galaxies with {\tt FLAG\_FIT}=0 or 2 when discussing $B/T$, and only galaxies with {\tt FLAG\_FIT}=0, 1, or 2 when discussing $r_{e,r}$.
For the kinematic parameters, $V$ and $\sigma$, we use the DAPall catalog (\cite{Westfall2019, Belfiore2019}).

First, we focus on the photometric morphology parameters.
From figure~\ref{fig;morph}~(a) and (b), we find that RGs have a very different morphology from nRQGs (the $p$-value of Kolmogorov-Smirnov test is $p\ll0.05$); RGs show more a disk-like morphology (lower $n_r$ and $B/T$) than nRQGs and a more elliptical-like morphology (higher $n_r$) than nRSFGs.
From figure~\ref{fig;morph}~(c), we find that RGs tend to have a larger radius than nRSFGs and a smaller radius than nRQGs. 
Similar to the $n_r$ and $B/T$ distributions, RGs have almost the same $R_{e,r}$ distribution as nRGVs.
Based on the above results, we conclude that the RGs have a photometric parameter distribution between nRQGs and nRSFGs, similar to nRGVs.

We can also see similar trends in the kinematic morphology parameters.
Figure~\ref{fig;morph}~(d) shows that RGs have an intermediate distribution of $V/\sigma$ between nRSFGs and nRQGs.
Particularly, RGs have a more rotation-supported morphology than nRQGs, even though the mass distribution of nRQGs is not significantly different from that of the RGs.
RGs also show a similar $V/\sigma$ distribution to nRGVs, but nRGVs have a slightly more rotational-supported than RGs.

\subsection{Environment}\label{s55}
To examine the environmental dependence of RGs, we cross-match our parent sample with the group catalog of \citet{Tempel2012} that applied a modified friends-of-friends method to the SDSS DR8, obtaining 7585 galaxies with environmental information.
We consider those with the richness $n_{\rm rich} =1 $ to be in the field or void environment, while those with $n_{\rm rich}>4$ to be in the cluster environment.

\begin{figure}
 \begin{center}
  \includegraphics[width=8cm]{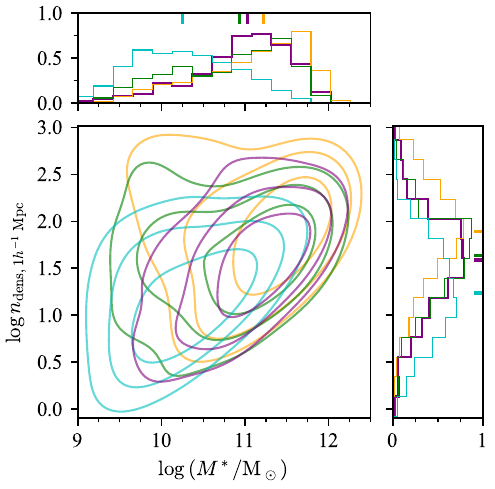}
 \end{center}
\caption{
Same as Figure~\ref{fig;morph}, but for stellar mass versus the scaled environmental density within $1h^{-1}~{\rm Mpc}$ from each galaxy. Individual galaxies are not plotted.

}
\label{fig;dens}
\end{figure}

Figure~\ref{fig;dens} compares among galaxy types the distribution of normalized environmental densities within $r=1h^{-1}~{\rm Mpc}$ that are calculated by \citet{Tempel2012} from galaxy luminosity density fields in the same manner as \citet{Liivamagi2012}.
We find that the density distribution of RGs is intermediate between those of QGs and SFGs, being similar to GVs.
This trend is similar to the morphology trend found in section~\ref{s54} and consistent with previous studies such as \citet{Schawinski2007, Thomas2010, Chauke2019} in that RGs tend to be in lower-density environments than QGs. 

\begin{figure}
 \begin{center}
  \includegraphics[width=8cm]{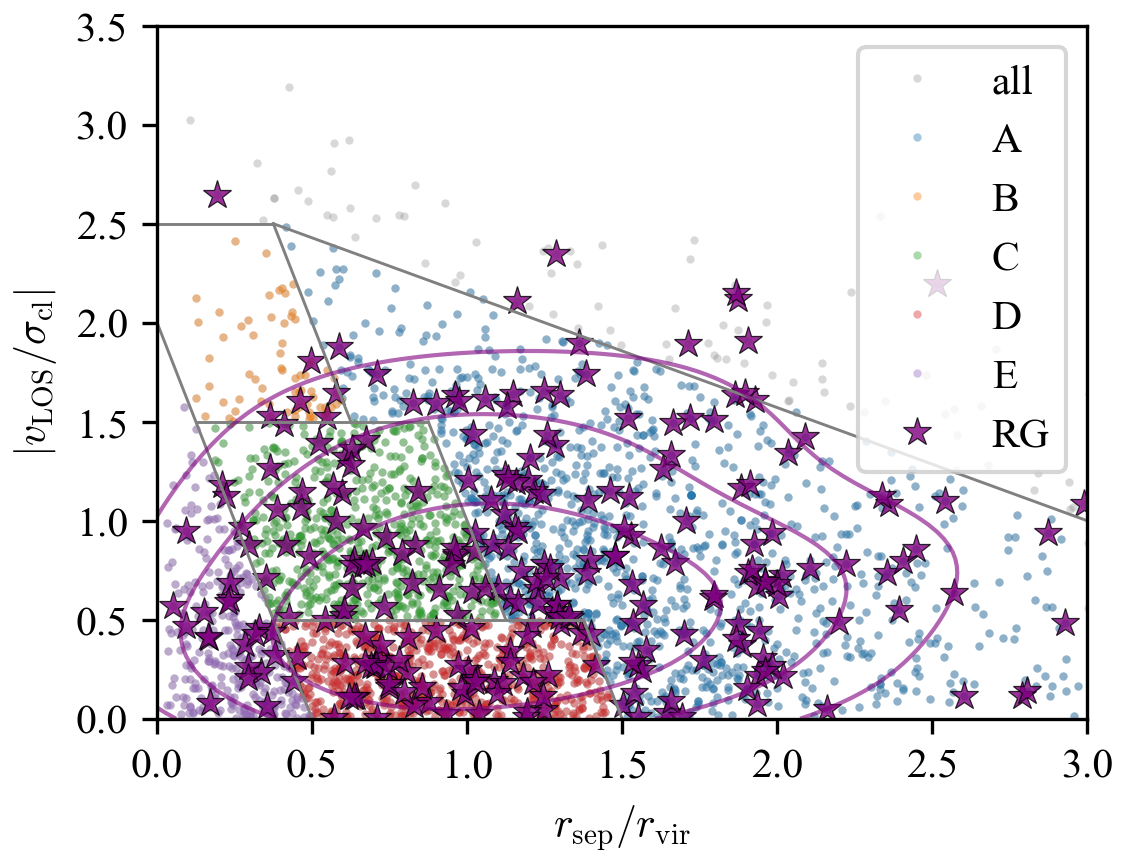} 
 \end{center}
\caption{
The phase-space diagram of our galaxies in the cluster environment.
Following \citet{Rhee2017}, we divide the cluster environment into five regions (A to E) whose boundaries are shown by solid gray lines.
Dots indicate all galaxies, color-coded according to the region they reside in. 
Purple stars are RGs, and purple contours are the 1 and $2\sigma$ levels of their distribution.
}
\label{fig;ps}
\end{figure}

Next, we draw a phase space diagram (\cite{Bertschinger1985}) for galaxies in the cluster environment ($n_{\rm rich}>4$), with the clustocentric radius $r_{\rm sep}$ normalized by virial radius $r_{\rm vir}$ being the horizontal axis and the clustocentric velocity in the line-of-sight direction $v_{\rm LOS}$ normalized the velocity dispersion in the cluster $\sigma_{\rm cl}$ being the vertical axis, using the group catalog by \citet{Tempel2012}.
Following \citet{Rhee2017}, we classify galaxies into five regions, A, B, C, D, and E, according to their positions on the phase-space diagram as shown in figure~\ref{fig;ps}.
\citet{Rhee2017} have reported that region A is dominated by interlopers and galaxies that have fallen into the cluster recently.
In contrast, region E is dominated by galaxies that have been in the cluster for a long time.

Table~\ref{tab;frac_env} shows $f_{\rm RG}$ for the five regions and the field.
We find that $f_{\rm RG}$ is higher in region A and the field than in regions B-D and E.
We also calculate $f_{\rm RG/nRQG}$, the ratio of the number of RGs to the number of nRQGs.
As RGs were QGs before rejuvenation, we consider $f_{\rm RG/nRQG}$ as an indicator of the fraction of QGs that caused rejuvenation, i.e., how likely rejuvenation occurs in QGs.
$f_{\rm RG/nRQG}$ is found to be highest in the field and decreases toward the cluster center in the order of regions A, B-D, and E.
This suggests that rejuvenation is more likely to occur in galaxies in the field and galaxies that have just entered a cluster than in galaxies that have been in a cluster for a long time.

\begin{table*}[]
\caption{The number of each type of galaxies for each environment. 
The error in the fraction only includes the Poisson error.
}
\label{tab;frac_env}
\begin{tabular}{ll | lllll | ll}\hline\hline
                             &       & \multicolumn{5}{c|}{Number}     & \multicolumn{2}{c}{Fraction} \\
                             &       & All & RG & nRSFG & nRGV & nRQG & $f_{\rm RG}$: RG/All       & $f_{\rm RG/nRQG}$; RG/nRQG       \\\hline
\multicolumn{2}{c|}{All Environment}              & 6743 & 731 & 2831 & 376 & 2805 & $10.8\pm0.4\%$ & $26.1\pm1.1\%$ \\
\multicolumn{2}{c|}{Field ($N_{\rm rich}=1$)} & 2063 & 221 & 1179 & 122 & 541 & $10.7\pm0.8\%$ & $40.9\pm3.3\%$ \\
\multirow{3}{*}{Cluster ($N_{\rm rich}>4$)} & A     & 1056 & 129 & 362 & 65 & 500 & $12.2\pm1.1\%$ & $25.8\pm2.5\%$ \\
                             & B-D   & 1294 & 102 & 326 & 69 & 797 & $7.9\pm0.8\%$ & $12.8\pm1.3\%$ \\
                             & E     & 272 & 20 & 38 & 13 & 201 & $7.4\pm1.7\%$ & $10.0\pm2.3\%$\\\hline
\end{tabular}
\end{table*}

%%%%%%%%%%%%%%%%% Discussion %%%%%%%%%%%%%%%%%%
\section{Discussion}\label{s6}
\subsection{RGs are important players in galaxy evolution}\label{dis:importance}
One of the key questions in the study of rejuvenation is the importance of RGs in galaxy evolution, especially how much of the total stellar mass was formed by rejuvenation and whether or not rejuvenation has a significant contribution to cosmic star formation.
\citet{Chauke2019} have reported that about 10\% of the total stellar mass of rejuvenated galaxies is formed in the rejuvenation event and that RGs contribute only about 0.3\% of the cosmic SFR density (CSFRD) over $z \sim 0.7-1.5$.
\citet{Tacchella2022} have reported that, at most, only about 10\% of the total mass is formed in rejuvenation events.

\begin{figure}
 \begin{center}
  \includegraphics[width=8cm]{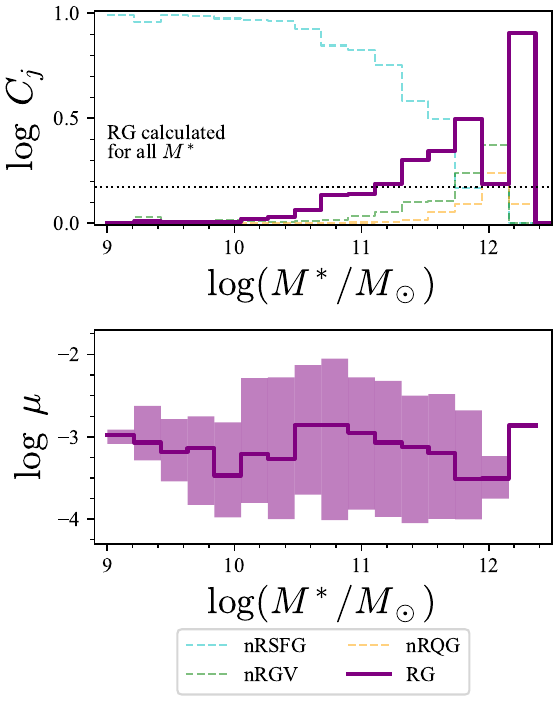} 
 \end{center}
\caption{
{\it Upper panel.} Distribution of $C_j$ as a function of $M^*$ for four types of galaxies coded by color. The horizontal dotted line indicates $C_{\rm RG}$, the contribution of RGs averaged over all masses. 
{\it Lower panel.} $\mu$ (purple solid line) and its $1\sigma$ confidence level (purple shaded area) as a function of $M^*$.
}
\label{fig;formed_mass}
\end{figure}

To address this question, first, we calculate $C_j$, the contribution of each current galaxy population $j$ ($j=$nRSFG, nRGV, nRQG, and RG) to the CSFRD in the last 100 Myr, the timescale of rejuvenation in our definition.
Here, $C_j$ is defined as the ratio of the total $M^*$ formed in $j$-type galaxies in the last 100 Myr to the total $M^*$ formed in all galaxies in the last 100 Myr,  
\begin{equation*}
    C_j = \frac{\sum_{i\in j}w_i M^*_{i,{\rm 100~Myr}}}{\sum_{i\in {\rm All}}w_i M^*_{i,{\rm 100~Myr}}},
\end{equation*}
where $M^*_{i,{\rm 100~Myr}}$ is the formed stellar mass in a galaxy $i$ in the last 100 Myr and $w_i$ is the volume weight of this galaxy given in the MaNGA catalog.
The mass dependence of $C_j$ is shown in the upper figure~\ref{fig;formed_mass}.
The sum of $C_{\rm RG}$ for all $M^*$ bins is about $17\%$, meaning that RGs contribute about $17\%$ to the CSFRD in the recent 100 Myr, about two orders of magnitude larger than the value obtained by \citet{Chauke2019} for rejuvenated galaxies at $z\sim0.8$.
If the detail of rejuvenation events is similar at $z\sim1$ and $z\sim0$, this result implies that rejuvenation events have higher importance at lower $z$.
However, this difference may be explained by the difference in the definition of rejuvenation and the difference in the parent sample (they search for RGs only from QGs).
It may also be attributed to the $z$ dependence of the CSFRD because the CSFRD at $z\sim1$ is about one order of magnitude higher than at $z\sim0$ (e.g., \cite{Madau2014}).
Note also that we cannot rule out the possibility that $C_j$ is overestimated or underestimated in this study due to the contamination or completeness discussed in section~\ref{s43}.

Next, we calculate $\mu$, the ratio of the formed stellar mass in the last 100 Myr, $M^*_{\rm 100~Myr}$, to the total stellar mass $M^*$ for each RG (same as the discussion in section~\ref{s42}).
The lower panel of figure~\ref{fig;formed_mass} shows $\mu$ as a function of $M^*$.
We find that rejuvenation events increase $M^*$ only by a few percent at most, with $\sim 0.1\%$ on average.
This result roughly agrees with \citet{Akhshik2021} reporting that $0.5\%$ of $M^*$ is formed in 100-Myr-timescale rejuvenation events at $z=1.883$.
This agreement may suggest that the detail of rejuvenation events does not change with redshift.

Note that the values of $\mu$ obtained here are consistent with the result for mock data presented in section~\ref{s42} that our method can select galaxies with $\mu\sim10^{-3}$ with high completeness.
This consistency suggests that our RG sample is not significantly contaminated.
On the other hand, we cannot rule out the possibility that $\mu \sim 10^{-3}$ is obtained just because our method fails to select RGs with other $\mu$ values.
To do a more solid statistical discussion, it will be important to evaluate contamination and completeness using a larger mock sample covering wider parameter ranges or maybe cosmological simulation data and to develop a more robust selection method.

The very small increase in $M^*$ by rejuvenation obtained here is primarily because the rejuvenation timescale, 100-Myr, is too short to increase $M^*$ significantly.
However, if galaxies experienced rejuvenation events many times in the past, rejuvenation may be an important channel of galaxy mass growth.
Although we cannot directly detect such past rejuvenation events for our sample because of insufficient time resolution of the reconstructed SFHs, we discuss the possibility of multiple rejuvenation events in section~\ref{dis:once_or_many}.

\subsection{Multiple rejuvenation scenario}\label{dis:once_or_many}
Whether rejuvenation can occur more than once in a galaxy's lifetime is still unclear.
\citet{Nelson2018} have found that $\sim10\%$ and $\sim1\%$ of massive passive galaxies in the TNG100 simulation have experienced one and more than one rejuvenation event, respectively.
In this subsection, we roughly estimate $N_{\rm rej}$, the number of times galaxies have experienced rejuvenation, using $f_{\rm RG}$ obtained in section \ref{s51}.

We simply assume that $f_{\rm RG}$ is $\sim8\%$ (the value obtained in section~\ref{s51}) over $z\sim0-1$. 
This assumption comes from the fact that previous studies of $z \lesssim 1$ galaxies have obtained similar $f_{\rm RG}$ values \citep{Tacchella2022,Chauke2019}.
We also estimate the timescale of rejuvenation $\tau_{\rm rej}$, the time from restarting star formation to stopping the resumed star formation, to be $\tau_{\rm rej} \left(10^8~{\rm yr}-10^{7.5}~{\rm yr}\right) / 0.21 \sim 320~{\rm Myr}$ because about 21\% of the RGs with $t_{\rm rej}=10^8~{\rm yr}$ returned to QGs at $10^{7.5}~{\rm yr}$.
Note that this calculation implicitly assumes a constant probability that the resumed star formation will end within a certain time.

With these assumption and estimate, the averaged $N_{\rm rej}$ for all galaxies between $z\sim0-1$ can be calculated as $N_{\rm rej} = \left(t|_{z=1}-t|_{z=0}\right)f_{\rm RG}/\tau_{\rm rej}\sim2.1$, suggesting that galaxies have experienced rejuvenation twice at $z \lesssim1$ on average.
Because $f_{\rm RG}$ increases with stellar mass as shown in figure~\ref{fig;SFRvsM}, massive galaxies can have larger $N_{\rm rej}$.
For example, the $N_{\rm rej}$ of $M^*>10^{11}M_\odot$ galaxies, with $f_{\rm RG}\sim18\%$, can be as high as $N_{\rm rej}\sim4.4$, meaning that such massive galaxies are likely to have experienced multiple rejuvenation events.

As shown in figure~\ref{fig;SFRvsM}, the path of a rejuvenation event in the SFR-$M^*$ plane is almost a vertical ascent.
Considering that the time scale of rejuvenation is $\tau_{\rm rej}\sim0.3~{\rm Gyr}$ and that rejuvenation may occur many times, essentially all rejuvenation events have to end up with re-quenching.
Therefore, we infer that massive galaxies are frequently moving up and down on the massive side of the SFR-$M^*$ plane.

To verify whether rejuvenation can occur many times, we use the reconstructed SFHs of current QGs to calculate the time from the last quenching $t_{\rm lq}$, the time since the galaxy last entered the QG regime from the SFG or GV regime.
In this verification, we consider two patterns of quenching, ``long quenching'' and ``mini quenching''.

In the ``long quenching'', it is assumed that galaxies never (at least for a longer time than $\tau_{\rm rej}$) rejuvenate once quenched.
This corresponds to the popular galaxy evolution scenario in that SFGs become QGs.
For simplicity, we assume that the probability of ``long quenching'' for SFGs does not change with cosmic time.
This assumption is reasonable because the cosmological evolution of quenching possibility can be ignored for our focusing timescale; e.g., if an observed object has $z\sim0.03$, even the second-to-last age bin only reaches about $z\sim0.2$, and the quenching rate at $z\sim0.2$ is about only two times higher than that at $z\sim0.03$ according to \citet{Peng2010}.

In contrast, ``mini quenching'' is temporary quenching that is terminated by rejuvenation.
Recently, \citet{Dome2023, Looser2023b} have reported that high-$z$ (low-mass) galaxies can experience mini-quenching.
\citet{Dome2023} have found the duration of a mini-quenching event to be $20-40$ Myr at $z\sim 7$, being comparable to the dynamical time of galaxies at this redshift.
Because the dynamical time scales as $t_{\rm dyn} \propto \rho^{-1/2} \propto \left(1+z\right)^{-3/2}$, we expect the duration of a mini-quenching event of local galaxies to be $\sim$400-900~Myr, which is indeed similar to the rejuvenation timescale $\tau_{\rm rej}$ of our RGs.
If galaxies can have multiple rejuvenation events, quenching other than the last mini quenching (i.e., the long quenching and the previous mini quenchings) will not be reflected in $t_{\rm lq}$ because $t_{\rm lq}$ only focuses on the last quenching event.
Therefore, the distribution of $t_{\rm lq}$ is expected to have an excess on a shorter time scale than $\tau_{\rm rej}$.

\begin{figure}
 \begin{center}
  \includegraphics[width=8cm]{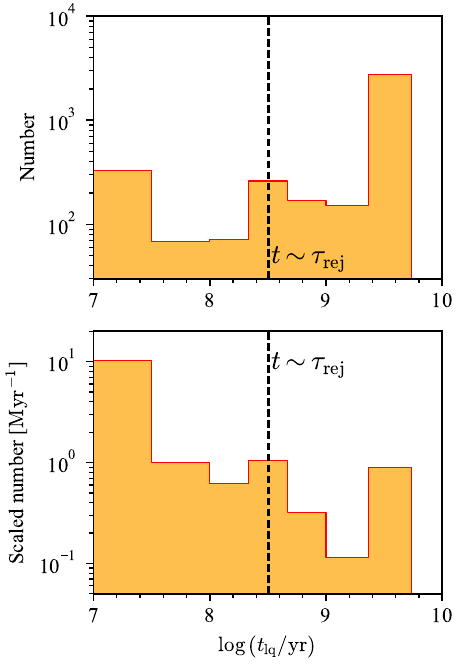} 
 \end{center}
\caption{
The distribution of $t_{\rm lq}$.
The upper panel plots the number of RGs falling into each $t_{\rm lq}$ bin while the lower panel plots a scaled number, i.e., the number divided by the bin width.
Vertical black dashed lines indicate $t_{\rm lq}=\tau_{\rm rej}$.
}
\label{fig;last_quenching}
\end{figure}

In the histogram of $t_{\rm lq}$ and the number normalized by the duration of those time bins shown in figure~\ref{fig;last_quenching}, we indeed find an excess with $t_{\rm lq}\lesssim\tau_{\rm rej}$.
This excess can be explained by considering that the information on mini-quenching that occurred at $t_l\gtrsim\tau_{\rm rej}$ is overwritten by rejuvenation and the next mini-quenching and hence that only mini-quenching at $t_l\lesssim\tau_{\rm rej}$ contributes to the histogram.
However, this interpretation is inconsistent with the fact that rejuvenation occurs at $z > 0$ as well, e.g., $z\sim0.8$ \citep{Chauke2019, Tacchella2022} and $z=1.88$ \citep{Akhshik2021}.
We conclude that galaxies, especially massive ones, have likely experienced multiple rejuvenation events.

\subsection{Why are they disk-like?}\label{s63}
As shown in figures~\ref{fig;morph}~(a) and (c), RGs have a more disk-like or rotational-supported morphology distribution than nRQGs, even though RGs were QGs before rejuvenation by definition.
In this section, we discuss the reason for this morphological difference between RGs and the nRQGs.
Here, we propose the following two hypotheses to explain the disk-like structure of RGs.

\begin{enumerate}
    \item {\bf Selective rejuvenation (SR)}\\
    As high as one-third $(=1259/3806)$ of the QGs in our sample have disk-like morphology, i.e., $V/\sigma>1$. 
    Furthermore, it is well known that some spiral galaxies are anemic spirals without star formation (e.g., \cite{vandenBergh1976,Shimakawa2022}).
    \citet{Fudamoto2022} have found the reddest spiral galaxies at $z\sim1-3$ from JWST/NIRCam images and reported that one of them is well-described as a passive galaxy.
    If only disk-like QGs, or passive spiral galaxies, can rejuvenate, and if their morphology does not change during rejuvenation, then RGs will have a disk-like morphology.
    This is ``selective'' rejuvenation, where only disk-like QGs can rejuvenate.
    
    \item {\bf Disk-forming rejuvenation (DR)}\\
    The disk-like morphology distribution can be explained if elliptical-like QGs form a disk during rejuvenation.
    This scenario does not prohibit the presence of RGs that evolved from disk-like QGs.
\end{enumerate}

\noindent 
We argue that the DR scenario is unlikely because of two serious problems.
First, it is extremely difficult to change morphology in a short period of $\tau_{\rm rej}\sim320~{\rm Myr}$.
Although \citet{Diaz2018} claim that elliptical galaxies can form a disk by merging with gas-rich satellite galaxies, it takes a Gyr time scale, much longer than $\tau_{\rm rej}$.
Besides, as shown in figure~\ref{fig;SFRvsM}, galaxies increase the stellar mass only by $\sim 0.1\%$ in a single rejuvenation event.
It is challenging to make a noticeable disk with such a small amount of mass increase.
Second, there are significant differences in morphologies between RGs and nRQGs (figure~\ref{fig;morph}).
These differences, especially in kinetics morphology $V/\sigma$, may not be explained by simple star formation in the outer part of galaxies due to rejuvenation.
For these reasons, we conclude that the DR scenario is unlikely.

We then test the SR scenario using the $M^*$ and $n_{{\rm dens}, 1h^{-1}~{\rm Mpc}}$ plane.
As shown in figure~\ref{fig;SFRvsM}, $M^*$ does not change significantly before and after rejuvenation.
Because $n_{{\rm dens}, 1h^{-1}~{\rm Mpc}}$ should be almost unchanged before and after rejuvenation, the spatial distribution of RGs and their progenitors in this plane should be the same.
By comparing the distribution of RGs, disk-like nRQGs, and elliptical nRQGs, we can test the SR scenario.

In the SR scenario, only disk-like QGs are the progenitors of RGs.
Because not all disk-like QGs may be able to rejuvenate, the distribution of disk-like QGs does not necessarily have to be the same as of RGs.
However, the distribution of disk-like QGs must cover that of RGs.
In contrast, elliptical QGs can take a significantly different distribution from RGs.

\begin{figure}
 \begin{center}
  \includegraphics[width=8cm]{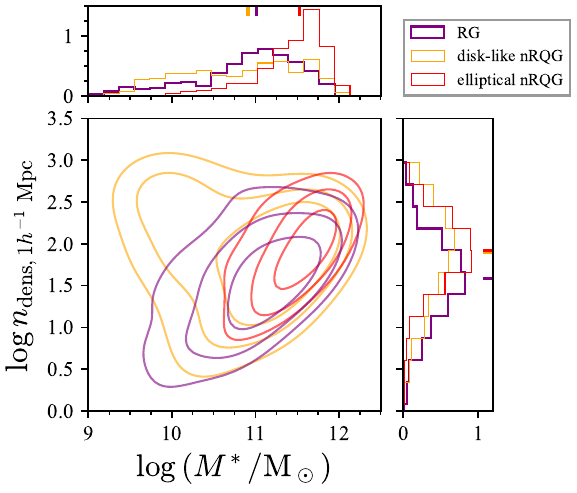} 
 \end{center}
\caption{
Same as Figure~\ref{fig;dens}, but for RGs (purple), disk-like nRQGs (orange), and elliptical nRQGs (red).
}
\label{fig;match}
\end{figure}

Figure~\ref{fig;match} shows the $M^*$ and $n_{{\rm dens}, 1h^{-1}~{\rm Mpc}}$ distributions of RGs, disk-like nRQGs, and elliptical nRQGs. 
Here, we define galaxies with the same $V/\sigma$ distribution with RGs as disk-like nRQGs and construct a sample of disk-like nRQGs by selecting an nRQG with the nearest $V/\sigma$ for each RG.
On the other hand, we define galaxies with $V/\sigma<0.5$ as elliptical nRQGs.
We find that 
all RGs are within the area where disk-like nRQGs are found, while a significant fraction of RGs are outside the area where elliptical nRQGs are found.
This result is consistent with the SR scenario.
Note that this result does not rule out the existence of RGs that evolved from elliptical-like QGs.
An interesting feature in this figure is that the distribution of disk-like nRQGs has a secondary peak around $\log\left(M^*/M_\odot\right)\sim 10$ and $\log\left(n_{{\rm dens}, 1h^{-1}~{\rm Mpc}}\right) \sim 2.5$ where no RGs exist. 
This may imply that low-mass sources in high-density regions cannot rejuvenate, consistent with the discussion in section~\ref{s55}.
In a subsequent paper, we will further discuss the mechanism of rejuvenation, including the conditions in which rejuvenation occurs.

Finally, we propose further tests for the two scenarios focusing on the mechanisms expected in these scenarios.
One possible mechanism with the SR scenario is that the remaining gas in disk-like QGs is used for rejuvenation.
In this case, disk-like QGs must have a higher molecular gas fraction $f_{\rm gas}$ than normal QGs, and the star formation in RGs must be driven by an increase in star formation efficiency (SFE).
Another possible mechanism in the SR scenario is that gas-poor disk-like QGs obtain gas to restart star formation.
In this case, we expect lower or similar $f_{\rm gas}$ in disk-like QGs. 
In any case, the SR scenario can be tested using a combination of $f_{\rm gas}$ and SFE data.
Additionally, the two scenarios can be verified by analyzing the spatially resolved SFH and kinetics of gases in RGs; e.g., evidence of gas inflows suggests the DR scenario.
Thus, a specially-resolved discussion of $f_{\rm gas}$, SFE, and kinetics using radio observation like ALMA and IFU data from MaNGA will help us constrain the mechanism and verify the SR and DR scenarios.
These hypotheses can also be tested by using simulation data to examine the progenitors of the RGs.

\subsection{The contribution of AGN}
In this study, we exclude 520 objects in section~\ref{s22} as AGN or merging galaxies.
This number amounts to only $6\%$ of the size of the parent sample and also smaller than the number of selected RGs.
Therefore, AGN activities as detected in MaNGA data are unlikely to be the primary driver of rejuvenation.
Note, however, that the parent sample may have some undetected AGNs and merging sources.
It is also possible that past AGN activities or mergers that are not detected from current observations could trigger rejuvenation because rejuvenation started $10^{7.5}~{\rm yr}$ or $10^8~{\rm yr}$ ago.

%%%%%%%%%%%%%%%%% Conclusion %%%%%%%%%%%%%%%%%%
\section{Conclusion}\label{s7}
In this paper, we have presented the {\sc Hinotori} project that aims at understanding the nature of RGs using the SFH.
As the first step of this project, we have performed a flexible SED fitting analysis using {\sc Prospector} on 8857 MaNGA galaxies with photometric and spectroscopic data and constructed a sample of 1071 RGs, the largest ever SFH-selected RG sample.
Our main results are summarized as follows:
\begin{itemize}
    \item Tests with mock data show that our selection method can select $f\sim0.001$ RGs with high completeness ($\sim67\%$).
    \item The RG fraction is $f_{\rm RG}=8.1\%$ after volume-weight correction.
    The selected RGs contribute $\sim 20\%$ of the CSFRD in the recent 100~Myr.
    \item From a rough estimate of $N_{\rm rej}$, we find that single galaxies can rejuvenate multiple times. 
    Multiple rejuvenations may significantly contribute to the mass assembly of galaxies, while a single rejuvenation increases the stellar mass only $\sim 0.1\%$.
    \item RGs have more disk-like morphology than nRQGs, suggesting the ``selective rejuvenation'' scenario where disk-like QGs rejuvenate more likely than elliptical QGs.
\end{itemize}

The next goal of {\sc Hinotori} is to specify the mechanism of rejuvenation.
We plan to analyze the IFU data of the selected RGs in the future.
Using simulation data is also an effective way to constrain the rejuvenation mechanism.
Furthermore, radio observations of molecular gas with, e.g., ALMA will allow us to discuss the star formation activity in RGs in more detail.
We also plan to examine the relationship between AGN activity and rejuvenation (c.f., \cite{Martin-Navarro2022}).

Recent studies have found high-$z$ low-$M^*$ QGs ($M^*\sim5\times10^8 M_\odot$ at $z\sim7.3$, \cite{Looser2023a}) and post-starburst galaxies ($M^*\sim4\times10^7 M_\odot$ at $z\sim5.2$, \cite{Strait2023}) in JWST data.
These findings may imply that high-$z$ low-mass QGs evolve into more massive galaxies at lower $z$ through repeated rejuvenation events (c.f., \cite{Looser2023a, Strait2023, Dome2023}).
To evaluate the importance and the role of rejuvenation in galaxy evolution, we have to explore RGs in a wide redshift range.
Besides, the $z$-dependence of $f_{\rm RG}$ will place stronger constraints on the possibility of multiple rejuvenations.
High-$z$ RG exploration needs a deep infrared spectroscopic survey with such as JWST/NIRSpec, TAO/SWIMS, Roman/WFI, and Euclid/NISP.
The development of a low-cost RG selection method (e.g., \cite{Zhang2022}) is another effective way to select RG candidates for follow-up spectroscopic observation and advance RG research.
As our results with mock data suggest, the outshining problem is serious when we reconstruct the SFHs of galaxies with strong rejuvenation.
Spatial resolved (or pixel-by-pixel) SED fitting may be able to solve this problem as \citet{Sorba2018} and \citet{Gimenez-Arteaga2023} suggested.

%%%%%%%%%%%%%%%%% Acknowledgement %%%%%%%%%%%%%%%%%%
\begin{ack}
% Hirashima & Iwasaki
We thank Keiya Hirashima and Kazunari Iwasaki for giving us useful advice for using a high-performance computing system.

% UGRASP and Cambridge
This work is partly supported by the 2022 Undergraduate Research Abroad in Science Program of School of Science, the University of Tokyo.
We thank Roberto Maiolino and other members of the Galaxy Formation and Evolution Group at Kavli Institute for Cosmology, Cambridge (KICC) for fruitful discussion in group meetings and with a cup of coffee.

% computation
This research is partially supported by Initiative on Promotion of Supercomputing for Young or Women Researchers, Supercomputing Division, Information Technology Center, The University of Tokyo. 
Numerical computations were in part carried out on Small Parallel Computers at Center for Computational Astrophysics, the National Astronomical Observatory of Japan, iDark cluster at Kavli Institute for the Physics and Mathematics of the Universe, The University of Tokyo, and the Fujitsu PRIMERGY CX400M1/CX2550M5 (Oakbridge-CX) at the Information Technology Center, The University of Tokyo. 

% Kakenhi, Grants, etc.
TT is supported by Forefront Physics and Mathematics Program to Drive Transformation (FoPM), a World-leading Innovative Graduate Study (WINGS) Program, the University of Tokyo.
H.Yesuf was supported by JSPS KAKENHI Grant Number JP22K14072 and the Research Fund for International Young Scientists of NSFC (11950410492).

% MaNGA https://www.sdss4.org/collaboration/#sdss4acknowledgement
% https://www.sdss4.org/collaboration/citing-sdss/
Funding for the Sloan Digital Sky Survey IV has been provided by the Alfred P. Sloan Foundation, the U.S. Department of Energy Office of Science, and the Participating Institutions. SDSS acknowledges support and resources from the Center for High-Performance Computing at the University of Utah. The SDSS web site is www.sdss4.org.

SDSS is managed by the Astrophysical Research Consortium for the Participating Institutions of the SDSS Collaboration including the Brazilian Participation Group, the Carnegie Institution for Science, Carnegie Mellon University, Center for Astrophysics | Harvard \& Smithsonian (CfA), the Chilean Participation Group, the French Participation Group, Instituto de Astrofísica de Canarias, The Johns Hopkins University, Kavli Institute for the Physics and Mathematics of the Universe (IPMU) / University of Tokyo, the Korean Participation Group, Lawrence Berkeley National Laboratory, Leibniz Institut für Astrophysik Potsdam (AIP), Max-Planck-Institut für Astronomie (MPIA Heidelberg), Max-Planck-Institut für Astrophysik (MPA Garching), Max-Planck-Institut für Extraterrestrische Physik (MPE), National Astronomical Observatories of China, New Mexico State University, New York University, University of Notre Dame, Observatório Nacional / MCTI, The Ohio State University, Pennsylvania State University, Shanghai Astronomical Observatory, United Kingdom Participation Group, Universidad Nacional Autónoma de México, University of Arizona, University of Colorado Boulder, University of Oxford, University of Portsmouth, University of Utah, University of Virginia, University of Washington, University of Wisconsin, Vanderbilt University, and Yale University.

%HSC
The Hyper Suprime-Cam (HSC) collaboration includes the astronomical communities of Japan and Taiwan, and Princeton University. The HSC instrumentation and software were developed by the National Astronomical Observatory of Japan (NAOJ), the Kavli Institute for the Physics and Mathematics of the Universe (Kavli IPMU), the University of Tokyo, the High Energy Accelerator Research Organization (KEK), the Academia Sinica Institute for Astronomy and Astrophysics in Taiwan (ASIAA), and Princeton University. Funding was contributed by the FIRST program from the Japanese Cabinet Office, the Ministry of Education, Culture, Sports, Science and Technology (MEXT), the Japan Society for the Promotion of Science (JSPS), Japan Science and Technology Agency (JST), the Toray Science Foundation, NAOJ, Kavli IPMU, KEK, ASIAA, and Princeton University. 

This paper is based on data collected at the Subaru Telescope and retrieved from the HSC data archive system, which is operated by the Subaru Telescope and Astronomy Data Center (ADC) at NAOJ. Data analysis was in part carried out with the cooperation of Center for Computational Astrophysics (CfCA), NAOJ. We are honored and grateful for the opportunity of observing the Universe from Maunakea, which has the cultural, historical and natural significance in Hawaii. 
\end{ack}

%%%%%%%%%%%%%%%%% Appendix %%%%%%%%%%%%%%%%%%
\appendix 
\section*{Comparison of {\sc Prospector}'s outputs with existing SED fitting results}
Several studies have estimated galaxy parameters for the MaNGA sample with SED fitting, although their results cannot be used to select RGs because they are based on parametric SFHs.
In this subsection, to further demonstrate the validity of our SED fitting results with {\sc Prospector}, we compare the estimated basic parameters with existing SED fitting results in the Pipe3D DR17 catalog \citep{Sanchez2022}.
This catalog summarizes the results of a stellar population and ionized gas analysis of the MaNGA data set using the pyPipe3D pipeline \citep{Lacerda2022}.
Note that the true values of individual parameters, like assumed mock values in section \ref{s41}, are unavailable and hence that we cannot know which SED fitting results are closest to the truth.
However, we expect our results to be generally consistent with the values in the Pipe3D catalog.

\begin{figure*}
 \begin{center}
  \includegraphics[width=16cm]{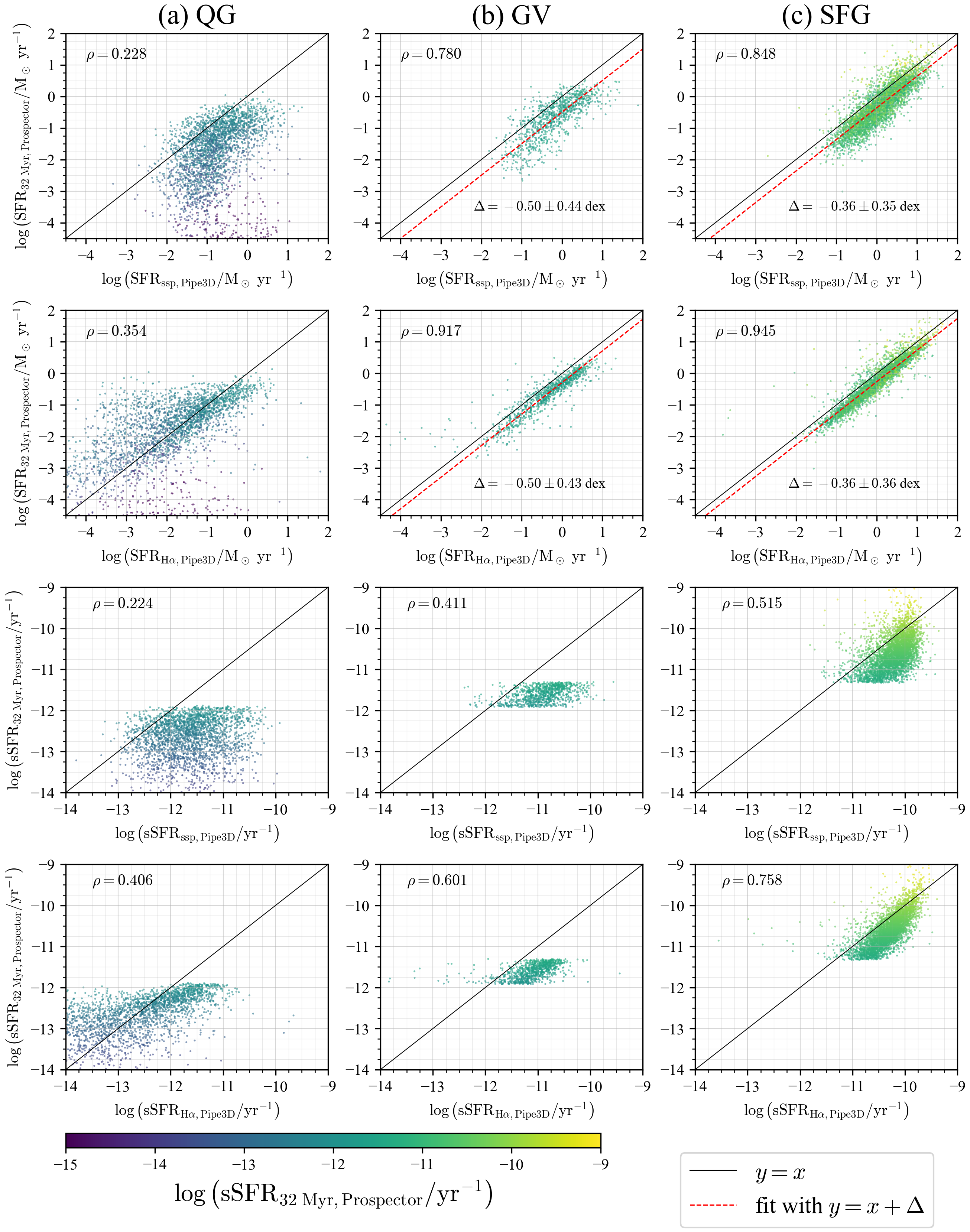} 
 \end{center}
\caption{
Comparison of Prospector-estimated SFRs and sSFRs with the Pipe3D catalog values.
In lines 1 and 2 (lnes 3 and 4), we compare estimated SFRs (sSFRs) in the most recent bin ($t<32~{\rm Myr}$) with Pipe3D SFRs (sSFRs) in recent $32~{\rm Myr}$ calculated from simple stellar population analysis and ${\rm H\alpha}$ fluxes, respectively.
Rows 1, 2, and 3 are for current QGs, GVs, and SFGs, respectively, classified with sSFRs in the most recent age bin estimated by {\sc Prospector}.
Spearman's correlation coefficient $\rho$ is shown in the upper left corner of each panel.
The solid black lines show the $y=x$ relation.
The dashed lines in $\rho>0.8$ panels show the results of $y=x+\Delta$ fit. 
}
\label{fig;pipe3d}
\end{figure*}

\begin{figure*}
 \begin{center}
  \includegraphics[width=16cm]{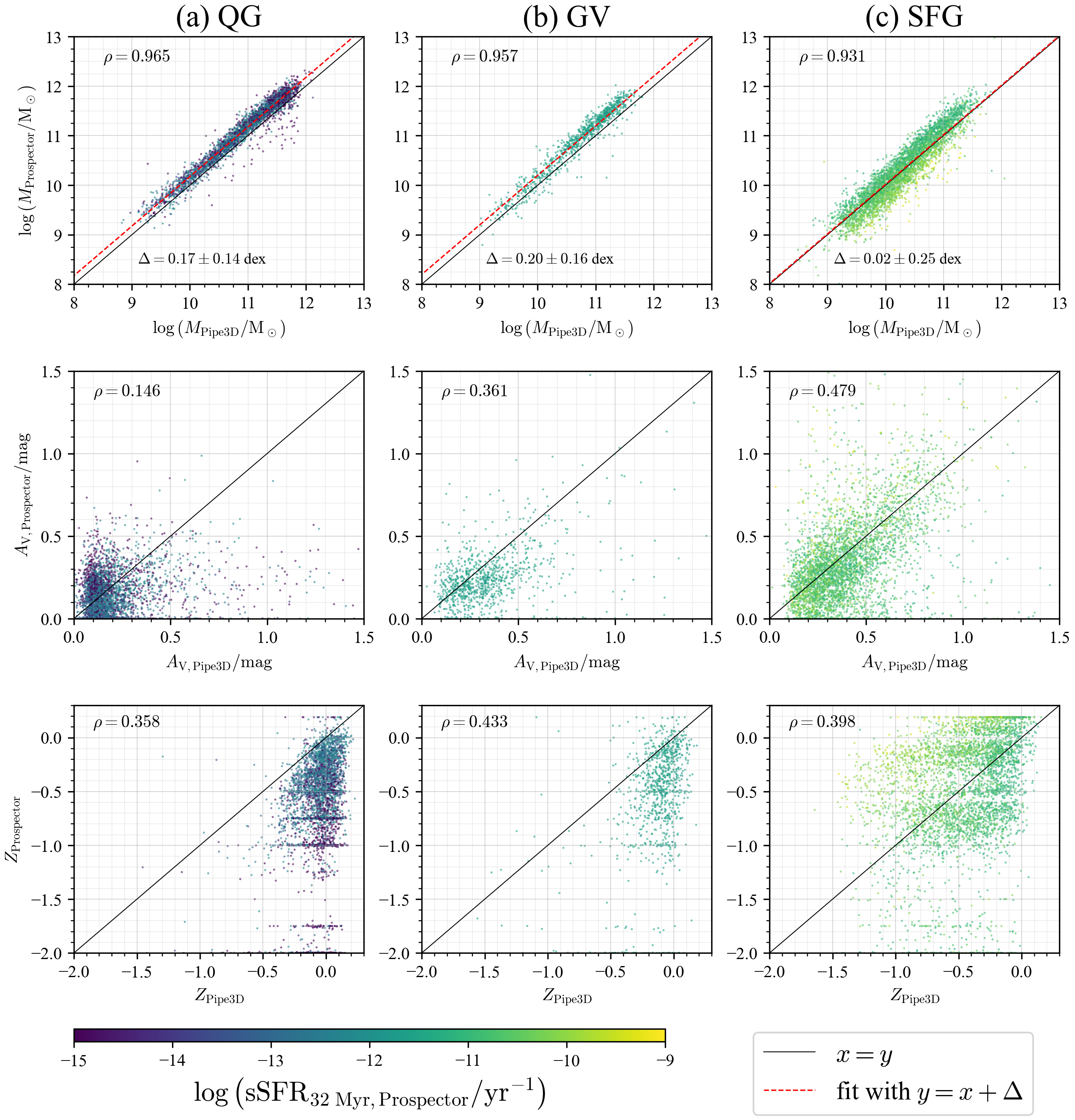} 
 \end{center}
\caption{
Same as figure~\ref{fig;pipe3d}, but in lines 1, 2, and 3, we compare estimated $M^*$, $A_{\rm V}$, and $Z$ with Pipe3D values.
}
\label{fig;pipe3d-2}
\end{figure*}

The Pipe3D catalog has two types of SFR; {\tt log\_SFR\_ssp} obtained from simple stellar population analysis and {\tt log\_SFR\_Ha} obtained from the Ha flux.
We use both SFRs in the Pipe3D catalog in the comparison.
Lines 1 and 2 of figure~\ref{fig;pipe3d} show that the SFRs estimated with {\sc Prospector} strongly correlate with those from the Pipe3D catalog for GVs and SFGs.
Indeed, Spearman's correlation coefficient calculated for all galaxies is $\rho\simeq0.75$ for SSP-based SFRs and $0.88$ for H$\alpha$-based ones.
We note, however, that the SFRs by {\sc Prospector} are systematically lower by about 0.6 dex.
In contrast, the correlation is very weak for QGs, especially those with ${\rm SFR}\lesssim 10^{-1}M_\odot~{\rm yr^{-1}}$.

Line 1 of figure~\ref{fig;pipe3d-2} shows that 
the stellar masses by {\sc Prospector} strongly correlate with those in the Pipe3D catalog ({\tt log\_Mass}), with a Spearman's correlation coefficient calculated for all types of galaxies of $\rho\simeq0.96$, although {\sc Prospector} tends to give slightly higher masses for QGs and GVs.
We also find that SFGs with higher sSFR in {\sc Prospector} tend to have lower $M^*$ than in the Pipe3D.

In our RG selection, we classify a galaxy into an SFG, GV, or QG at each of the eight age bins using the sSFR and $t_H$ of the bin.
Because the redshift range of galaxies in the parent sample is very narrow ($0.01\le z\le0.15$), the differences in $t_H$ at each age bin among the galaxies are less than $\sim 15\%$.
Therefore, comparing {\sc Prospector}'s sSFRs in the most recent age bin with the Pipe3D's values provides a sensitive test for the validity of our RG selection.

Lines 3 and 4 of figure~\ref{fig;pipe3d} compare sSFRs 
for each galaxy type.
For the whole sample, we obtain a high Spearman's correlation coefficient of $\rho\simeq0.84$ (vs. SSP-based sSFR) and $0.91$ (H$\alpha$-based sSFR).
However, as seen from the figure, the correlation in a given galaxy type is not so strong, being weaker than that for SFR and $M^*$.
This is mainly because the classification of SFGs, GVs, and QGs is based on sSFR estimated by {\sc Prospector}; thus, the distribution in each panel has a cutoff at an sSFR.

Line 2 of figure~\ref{fig;pipe3d-2} shows that the $A_{\rm V}$ estimated with {\sc Prospector} are weakly correlated with those from the Pipe3D catalog ({\tt Av\_ssp\_Re}) with a Spearman's correlation coefficient (for all galaxies) of $\rho\simeq0.48$.
The correlation is weakest for QGs.
Similarly, we find  a very weak correlation for $Z$ ({\tt ZH\_LW\_Re\_fit} in Pipe3D) in line 3 of figure~\ref{fig;pipe3d-2}.

In summary, the values of SFR, $M^*$, and sSFR by {\sc Prospector} correlate strongly with those from the Pipe3D catalog despite the presence of systematic offsets.
Therefore, we conclude that  our results are reliable enough to select RGs.
The criteria for RG selection are based solely on {\sc Prospector}' outputs.
This means that the systematic offsets from  Pipe3D's values will not significantly affect our classification.

%%%%%%%%%%%%%%%%% Reference %%%%%%%%%%%%%%%%%%

\end{document}